\documentclass{aa}  

\usepackage{graphicx}
\usepackage{rotating}
\usepackage{txfonts}
\usepackage{natbib, twoopt}
\bibpunct{(}{)}{;}{a}{}{,}
\usepackage{lipsum}
\usepackage{hyperref}
\usepackage{booktabs}
\usepackage[font=small]{caption}
\hypersetup{
        colorlinks = true,
        linkcolor = blue,
        citecolor = blue,
        filecolor = magenta,
        urlcolor = blue}
\usepackage{subcaption}        
\usepackage{lscape}             
\usepackage{placeins} 
\usepackage{dblfloatfix}
\usepackage{hyperref}

\begin{document}

\authorrunning{Covella et al.}
\titlerunning{A Walk on the Retrograde Side: III}

   \title{A Walk on the Retrograde Side (WRS) project:}
   \subtitle{III. High-resolution chemical makeup of the disrupted metal-poor stellar cluster ED-2}

    \author{S. Covella\inst{1}\fnmsep\inst{2},
    A. Mucciarelli\inst{1}\fnmsep\inst{2},
    D. Massari\inst{2},
    E. Ceccarelli\inst{3},
    M. Bellazzini\inst{2}, 
    E. Dodd\inst{4}}

    \institute{Dipartimento di Fisica e Astronomia "Augusto Righi", Università degli Studi di Bologna, Via Gobetti 93/2, I-40129 Bologna, Italy\\
    \email{sara.covella2@unibo.it}
    \and
    INAF, Osservatorio di Astrofisica e Scienza dello Spazio di Bologna, Via Gobetti 93/3, I-40129 Bologna, Italy
    \and
    Kapteyn Astronomical Institute, University of Groningen, Landleven 12, 9747 AD Groningen, The Netherlands
    \and
    Institute for Computational Cosmology \& Centre for Extragalactic Astronomy, Department of Physics, Durham University, South Road, Durham, DH1 3LE, UK}

  \abstract
  {In the context of the Milky Way's assembly history, ED-2 represents a particularly intriguing substructure, as the nature of its progenitor, whether a star cluster or a dwarf galaxy, remains uncertain. Previous studies have addressed this question by investigating the spread of the metallicity distribution function (MDF), one of the main chemical differences between these two types of systems. These studies found a narrow MDF consistent with a star cluster origin, although their conclusions were limited by the spectral resolution and sample size of the available data. In this project, we aim to constrain the nature of ED-2 by analysing high-resolution ($\rm R \approx 40000$) and high signal-to-noise ratio (S/N $\geq 30$) spectra for a sample of 23 stars. We perform a Bayesian analysis of the MDF using a Markov Chain Monte Carlo approach to infer the intrinsic metallicity dispersion, and investigate the abundance patterns of light elements to search for the variations typically observed in globular clusters. We find a mean metallicity of [Fe/H] $= -2.29^{+0.03}_{-0.02}$ dex and an intrinsic metallicity dispersion consistent with zero within the uncertainties. Similarly, the abundances of Mg, Na, and Ca show no significant intrinsic dispersion. The chemical homogeneity of ED-2 supports a stellar cluster origin and is consistent with a metal-poor system accreted into the Milky Way halo. The lack of significant light-element abundance variations suggests a low original mass, insufficient to retain the gas ejected by low-energy stellar sources, pointing to ED-2 being an old, metal-poor, and particularly low-mass star cluster.}


   \maketitle
   \nolinenumbers

\section{Introduction}\label{intro}
Within the hierarchical framework of the $\Lambda$ Cold Dark Matter cosmology, the stellar halo of the Milky Way (MW) is expected to have formed through the continuous accretion and disruption of lower-mass systems, whose remnants preserve information on the Galaxy's past merger history (e.g. \citealt{White+78}, \citealt{White+91}, \citealt{Springel+05}). Identifying and characterising these accreted stellar populations is therefore essential for constraining the formation pathways of the MW (e.g. \citealt{Helmi2020}).\\
From the dynamical point of view, the debris of a disrupted satellite evolves within the Galactic potential, progressively becoming more spatially incoherent as a consequence of phase mixing. While tidal streams provide direct evidence of ongoing or relatively recent accretion events, older debris becomes increasingly difficult to identify as spatially coherent structures (\citealt{Helmi2018}, \citealt{Belokurov+18}). Nevertheless, stars originating from the same progenitor are expected to retain coherence in dynamical quantities, such as the integrals of motion (IoM), long after spatial signatures have disappeared (\citealt{Helmi2000}). Consequently, the distribution of stars in IoM space offers a powerful tool to recover ancient accretion events and to probe the hierarchical build-up of the Galactic halo (\citealt{Koppelman+19}, \citealt{Massari+19}, \citealt{Dodd+23}, \citealt{Naidu+20}, \citealt{Myeong+19}).\\
However, the interpretation of dynamical substructures is not always straightforward. Debris originating from a single progenitor may fragment into multiple overdensities in the IoM space, while unrelated accretion events can overlap, complicating the association between dynamical structures and their progenitors (\citealt{Koppelman+20}; \citealt{Belokurov+22}, \citealt{Amarante+22}, \citealt{Thomas+25}). Moreover, the dynamically identified structures may originate from a variety of progenitor systems, ranging from low-mass galaxies to smaller stellar systems, such as star clusters, which may have been fully or partially disrupted. Distinguishing between these two kinds of progenitor systems is crucial, as dwarf galaxies and star clusters contribute differently to the build-up of the galactic halo, with typical differences in their total masses spanning up to several orders of magnitude.\\ 
In this context, stellar chemical abundances provide an independent and complementary diagnostic, as they preserve the imprint of the star formation and chemical enrichment history of the parent system (\citealt{Freeman+02}). The joint chemo-dynamic approach has enabled a more robust identification of accreted stellar populations and a more complete reconstruction of the MW assembly history (\citealt{Gallart+05}; \citealt{Tolstoy+09}; \citealt{Nissen+10}; \citealt{Hayes+18}; \citealt{Mackereth+19}; \citealt{Hasselquist+21}; \citealt{Horta+21}; \citealt{Minelli+21}; \citealt{Mucciarelli+21b}; \citealt{Buder+22}; \citealt{Matsuno+22}; \citealt{Horta+23}; \citealt{Ceccarelli+24}; \citealt{Ceccarelli+25}; \citealt{Mori+26}). 
\\
\\
Among the stellar substructures identified in the galactic halo in recent years, this work focuses on ED-2, a dynamical structure discovered by \citealt{Dodd+23} using Gaia Data Release 3 (DR3, \citealt{GaiaCollaborationphotometry+23}) data through the application of a clustering algorithm (see \citealt{Ruiz-Lara+22}, \citealt{Lovdal+22}) in the three-dimensional space of IoM, defined by the orbital energy and the angular momentum components ($E, L_z, L_\perp$).\\
\begin{figure}[h]
  \centering
  \includegraphics[width=\columnwidth]{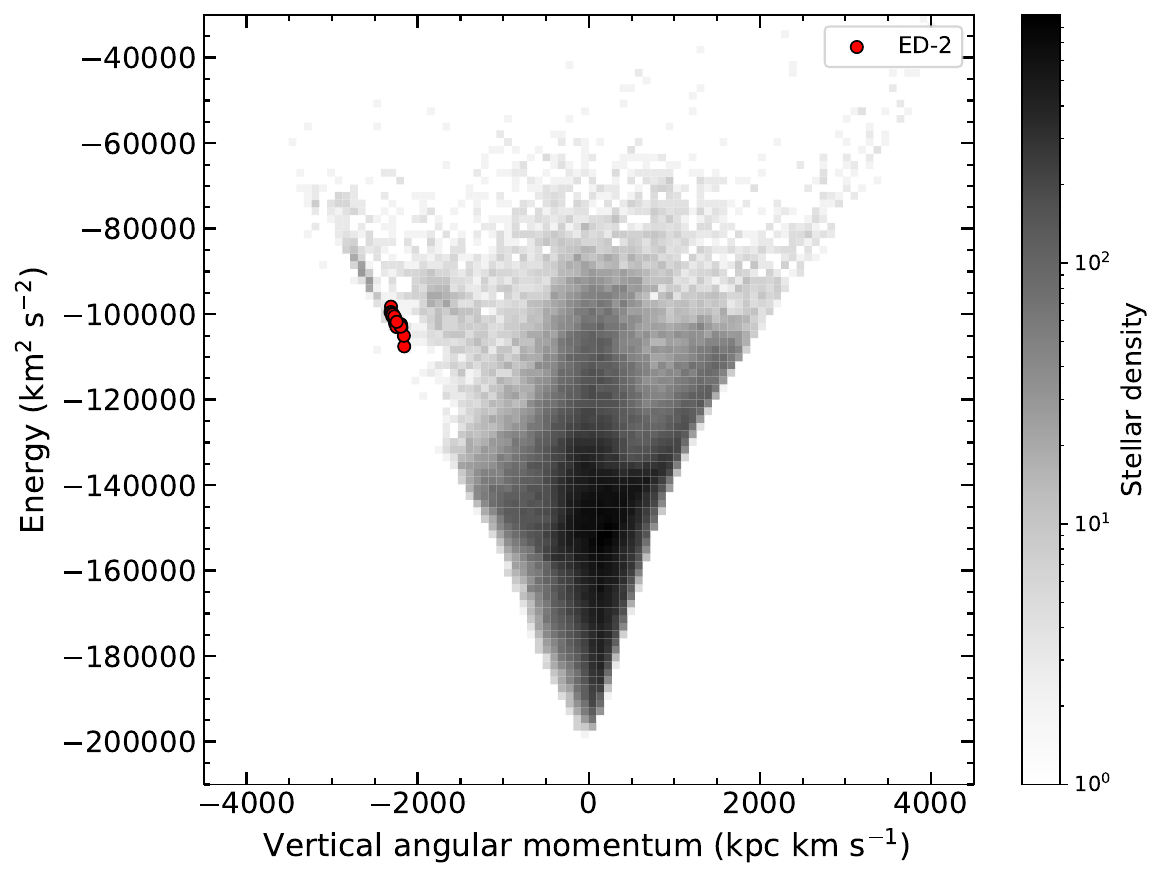}
  \caption{IoM space where the energy is plotted as a function of the vertical angular momentum. Among all the stars listed in the catalogue provided by \citealt{Dodd+23} represented in grey, the stars dynamically associated to ED-2 are represented in red.}
  \label{fig:IoM_space}
\end{figure}\\
As shown in Fig.\ref{fig:IoM_space}, ED-2 is characterised by a highly retrograde orbit and narrow distributions in both energy and angular momentum, as well as a compact velocity distribution (see \citealt{Dodd+25}). These properties, together with the spatial coherence observed in the $x$--$z$ plane, suggest that ED-2 has not yet undergone significant phase mixing and retains a coherent dynamical signature in both configuration and velocity space (\citealt{Balbinot+23}).\\
\\
The nature of the progenitor of ED-2 remains unclear. While halo substructures are often associated with the debris of disrupted dwarf galaxies, the photometric properties of ED-2 point towards a different scenario. In particular, the colour–magnitude diagram (CMD) of its member stars is consistent with a single-age stellar population  of 13.5 Gyr (\citealt{Balbinot+23}, \citealt{Dodd+25}), which is more commonly associated with star clusters.\\
To further investigate the nature of ED-2, subsequent spectroscopic studies have focused on the spread of its metallicity distribution function (MDF), which is the most fundamental chemical difference between star clusters and dwarf galaxies (\citealt{Willman+12}, \citealt{Cerny+26}, \citealt{Bellazzini+26}). The first spectroscopic analysis of ED-2 stars was presented by \citealt{Balbinot+23}, based on LAMOST DR18 low- and medium-resolution spectra for a sample of 21 stars and it suggested a relatively narrow MDF but with significant uncertainties. Subsequent works (see \citealt{Ceccarelli+24, Dodd+25}), using high-resolution spectra, confirmed a negligible metallicity spread, although the analysis was based on very limited samples (4 and 5 stars respectively). The combined limitations in spectral resolution and sample size have so far prevented a definitive determination of the nature of the ED-2 progenitor.\\
As part of the WRS project (\citealt{Ceccarelli+24}), which aims at chemically tagging retrograde stars in the Galactic halo through high-resolution spectroscopy, with the goal of disentangling accreted substructures and reconstructing the Milky Way's accretion history (\citealt{Ceccarelli+25}), in this work, we analysed high-resolution spectra of a sample of 23 stars dynamically associated to ED-2, enabling a more robust assessment of the nature of the ED-2 progenitor through the analysis of its chemical properties.\\
The paper is structured as follows. Section~\ref{Target} presents the target selection criteria and the high-resolution spectroscopic dataset. Section~\ref{params} describes the determination of the stellar atmospheric parameters, while Sect.~\ref{LOS} describes the measurement of the line-of-sight velocities. The methodology adopted to derive the chemical abundances is presented in Sect.~\ref{methods}. The resulting chemical abundance analysis is discussed in Sect.~\ref{results}. Finally, Sect.~\ref{disc_conc} summarises the main findings and presents our conclusions.\\

\section{Target selection and observations}\label{Target}
Our dataset consists of 23 stars (both dwarf and giants) selected as members of ED-2 according to the selection by \citealt{Balbinot+23} and in the range of Gaia G-band magnitude $10.8 \leq \rm G \leq 14.3$.\\
\\
The final sample includes proprietary and archival spectra for the 23 ED-2 stars, which were observed with the high-resolution spectrographs VLT/UVES (\citealt{Dekker+00}) and LBT/PEPSI (\citealt{Strassmeier+15}). All the information about the spectroscopic targets are listed in Tab.\ref{tab:abundances1}. In particular, 7 stars are observed with UVES, 15 stars are observed with PEPSI and one star has been observed with both spectrographs and is used to cross-validate the results from the two instruments. \\
Our sample includes all previously studied ED-2 members with available high-resolution spectroscopy, which are now re-analysed within a homogeneous framework. Specifically, our sample consists of 21 stars previously analysed by \citealt{Balbinot+23} using medium-resolution spectra ($\rm R\approx7500$). Some of these targets were also studied by \citealt{Ceccarelli+24} and \citealt{Dodd+25}, using spectra with a resolution of $\rm R=40000$. In addition, we include two stars, labelled ED2-18 and ED2-23, not included in previous works. ED2-18 is a candidate ED-2 star that has never been analysed before and here we have used proprietary UVES spectra in the configuration described below.\\
ED2-23, instead, is the visible companion of the black hole Gaia BH3. It was first analysed by \citealt{Panuzzo+24} and by \citealt{Balbinot+24} on the basis of archival UVES spectra acquired in Dichroic mode adopting the standard settings Dic 1 Blue Arm CD2 390 (326–454 nm) and Dic 1 Red Arm CD3 580 (476–684 nm) and with the 0.7” slit width, thus yielding a resolution of R$\approx$55000. Subsequently, \citealt{Hackshaw+25} obtained 44 hr of observations of this target over 11 nights in 2024 using the Tull Coudé spectrograph at the 2.7 m Harlan J. Smith Telescope. The observations were acquired with the TS23 setup and Slit 4, providing a spectral resolution of R$\sim$60000 and covering the wavelength range 3700–9000 Å. The combined spectrum reached a S/N$\sim$200, allowing for the detailed determination of elemental abundances, in particular of heavy r-process elements. More recently, this target was re-analysed by \citealt{Broeck+26} using UVES spectra acquired with dichroic \#1 and dichroic \#2 and the standard UVES settings 346 + 580 and 437 + 860, respectively, covering the optical range from 305 to 1040 nm. These observations were obtained with a slit width of 0.4–0.5'', yielding a spectral resolution of R = 65000.\\ 
\\
PEPSI observations were carried out with the CD3 grating for the Blue Arm, covering the wavelength range [4800, 5440] Å, and the CD4 and CD6 gratings for the Red Arm covering [5364, 6321] Å and [7410, 9140] Å respectively, with spectral resolution of 50000. The data were reduced using the PEPSI pipeline, including bias subtraction, flat-field correction, order extraction, wavelength calibration, and normalisation. In particular, sky subtraction was performed using dedicated sky exposures.\\
\\
UVES observations, instead, were performed in Dichroic mode adopting the standard settings Dic 1 Blue Arm CD2 390 covering the wavelength range [3600, 4800] Å and Dic 1 Red Arm CD3 564 covering the wavelength range [4800, 6600] Å and with the 1" $\times$ 12" slit, corresponding to a spectral resolution of 40000. The data reduction was carried out with the ESO UVES pipeline, following standard procedures. Sky subtraction was performed using the sky spectrum within the slit.\\
\\
\begin{figure}[!h]
  \centering
  \includegraphics[width=0.8\columnwidth]{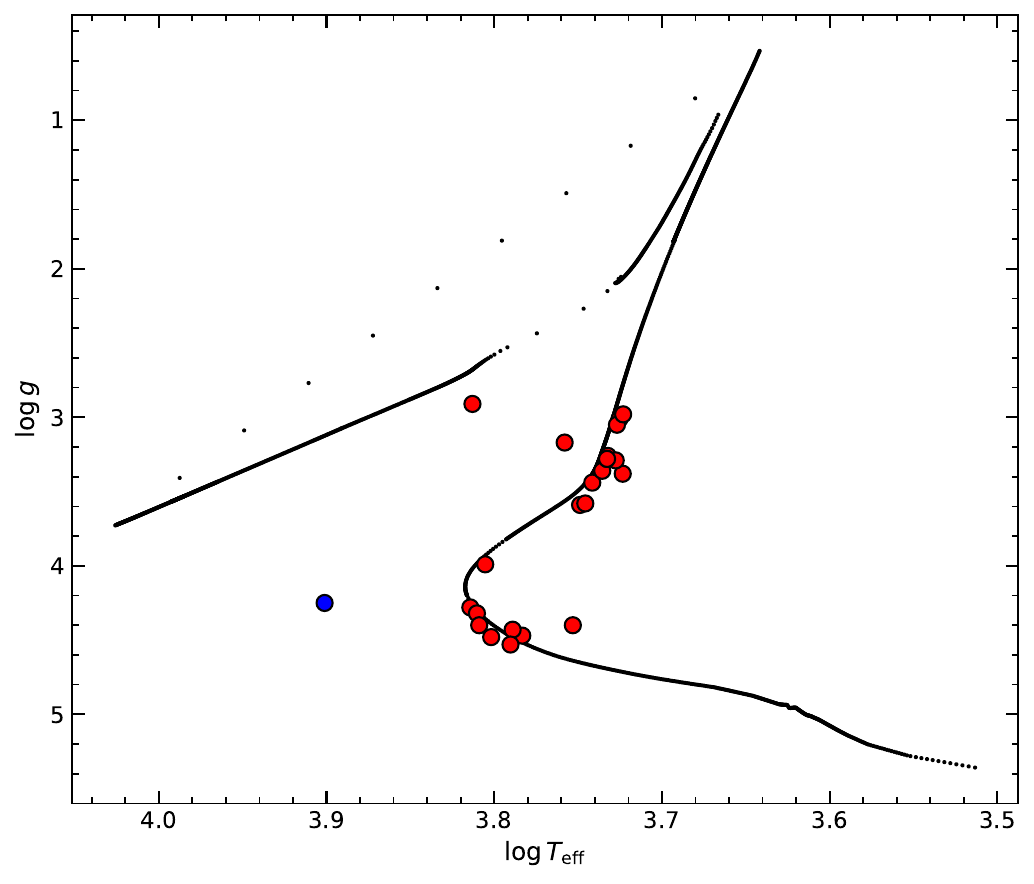}
  \caption{Kiel diagram for the spectroscopic targets (red circles). The black points are the BaSTI isochrone (\citealt{Pietrinferni+21}) with age of 13.5 Gyr, [Fe/H]=-2.44 dex and [$\alpha$/Fe] = +0.4 dex chemical mixture. The hottest star in our sample, which is most likely a blue straggler star, is here represented in blue.}
  \label{fig:CMD_Kiel}
\end{figure}

\section{Stellar parameters}\label{params}
We derived the stellar atmospheric parameters, namely the effective temperature and surface gravity, for all targets using the photometry provided by Data Release 3 (\citealt{GaiaCollaborationphotometry+23}) of the Gaia space mission (\citealt{Prusti16}), while the microturbulent velocity was determined through the optimisation procedure described in Sect.\ref{vt}.\\
\\
The effective temperatures ($\rm T_{eff}$) were obtained from the de-reddened $\rm (BP-RP)_0$ colour using the $\rm (BP-RP)_0-\rm T_{eff}$ relation by \citealt{Mucciarelli+21}.\\
To apply this relation, the observed $\rm (BP - RP)$ colour has been corrected for extinction using the iterative procedure described in \citealt{GaiaCollaboration2018}. Specifically, the colour excess $\rm E(B-V)$ adopted for each star is provided by
the online \texttt{G-TOMO} tool within the \texttt{EXPLORE} platform (\citealt{EXPLOREplatform}), which extrapolates the $\rm E(B-V)$ from the 3D reddening map (\citealt{Lallement+22}). To extract the $\rm E(B-V)$ from the 3D map we used the distance obtained from Gaia DR3 parallax, corrected using the zero-point correction by \citealt{Lindegren+21}.\\
In addition to this, the relation by \citealt{Mucciarelli+21} requires as input a value for the Fe abundance which is here taken to be [Fe/H] = $-2.44$ dex for all stars, consistent with previous high-resolution studies of ED-2 (\citealt{Ceccarelli+24}, \citealt{Dodd+25}).\\
Uncertainties on $\rm T_{eff}$ were estimated by summing in quadrature three sources of errors: (i) the intrinsic uncertainty of the calibration relation, taken to be 60–80 K for dwarf and giant stars respectively as quoted by \citealt{Mucciarelli+21}; (ii) the photometric uncertainty on the $\rm (BP-RP)$ observed colour derived as sum in quadrature of the uncertainties for the BP and RP magnitudes; and (iii) the uncertainty on the reddening, for which a conservative 20\% relative error on $\rm E(B-V)$ was assumed. Specifically, this last assumption accounts for the extrapolation of the extinction curves performed by the \texttt{G-TOMO} tool for targets located outside the limits of the reddening map.\\
\\
Surface gravities ($\log g$) were derived by estimating the stellar radius according to the Stefan-Boltzmann relation, where luminosities were computed from Gaia DR3 G-band magnitudes, parallaxes and computing the G-band bolometric corrections according to \citealt{Andrae18}.\\
Stellar masses, instead, were estimated from a BaSTI isochrone assuming [Fe/H] = $-2.44$ dex, an $\alpha$-enhanced ([$\alpha$/Fe] = +0.4 dex) composition and an age of 13.5 Gyr (\citealt{Dodd+25}), resulting in typical values of $\rm 0.72 \ M_\odot$ and $\rm 0.80 \ M_\odot$ for dwarf and giant stars, respectively. The uncertainties on $\log g$ are derived by propagating the uncertainties on the effective temperature, photometry, and distance, yielding an uncertainty of $\sim0.1$ dex, comparable to the value reported by \citealt{Ceccarelli+24}.\\
\\
The derived $\rm T_{eff}$ and $\log g$ of the spectroscopic targets are shown in Fig.\ref{fig:CMD_Kiel}, superimposed with the BaSTI isochrone with age 13.5 Gyr, [Fe/H]=-2.44 dex and [$\alpha$/Fe] = +0.4 dex chemical mixture.

The Kiel diagram reveals the presence of a hot star in our sample (ED2-18 represented in blue in Fig.\ref{fig:CMD_Kiel}), with an effective temperature of $\rm T_{eff} = 7964$ K and a $\log g = 4.25$ dex. The comparison of its position in the Kiel diagram with an appropriate BaSTI zero-age horizontal branch model computed for [Fe/H] = -2.44 dex and [$\alpha$/Fe] = +0.4 dex chemical mixture indicates that this star might be a blue straggler. Its spectrum is dominated by Balmer lines, preventing a reliable determination of chemical abundances. Indeed, for this star we were only able to derive the line-of-sight velocity via cross-correlation technique and to place an upper limit on its metallicity.\\

\section{Line of sight velocities}\label{LOS}
\begin{figure}[h]
  \centering
  \includegraphics[width=0.9\columnwidth]{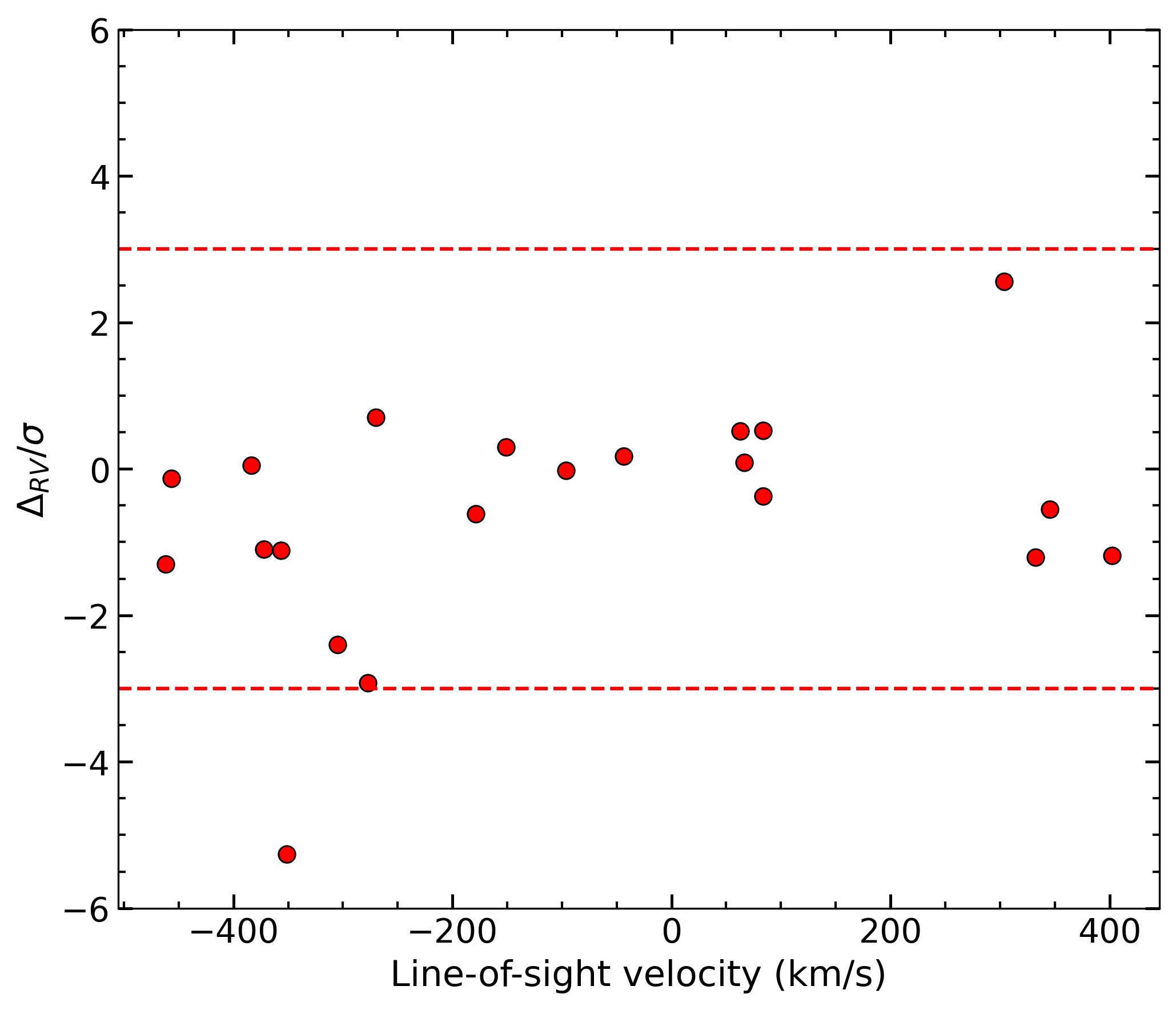}
  \caption{Normalised difference of the $\rm V_{LOS}$ derived in this study and by Gaia as a function of the spectroscopic $\rm V_{LOS}$. From this plot we checked the presence of binary stars within our sample of targets. The only star consistent with this nature is ED2-23, the Gaia BH3 companion star.}
  \label{fig:RV}
\end{figure}

Line-of-sight velocities ($\rm V_{LOS}$) were measured using the cross-correlation technique, by comparing observed and synthetic spectra computed with \texttt{SYNTHE} (\citealt{Kurucz+05}). The templates were generated adopting the stellar parameters of each target (see Sect.~\ref{params}) and convoluted with appropriate Gaussian profiles in order to reproduce the observed line broadening. For each star, the $\rm V_{LOS}$ was determined from the peak of the cross-correlation function and applied to correct the observed spectrum to the rest frame.\\
\\
To assess the accuracy of the measured $\rm V_{LOS}$, we compared our measurements with those provided by \citealt{GaiaCollaborationphotometry+23}. The agreement was quantified by computing, for each star, the normalised difference that is defined as follows:
\begin{equation}\label{norm_diff}
    \frac{\Delta \rm V_{LOS}}{\sigma}=\frac{\rm V_{LOS}^{ours}-\rm V_{LOS}^{Gaia}}{\sqrt{\sigma_{\rm V_{LOS}^{ours}}^2+\sigma_{\rm V_{LOS}^{Gaia}}^2}}
\end{equation}
where $\rm V_{LOS}^{ours}$ and $\rm V_{LOS}^{Gaia}$ denote the velocities derived in this work and those from \citealt{GaiaCollaborationphotometry+23}, respectively. The results of the normalised difference are shown in Fig.\ref{fig:RV}, where the normalised difference is plotted as a function of $\rm V_{LOS}^{ours}$. All stars but one are statistically consistent with Gaia values within the 3-$\sigma$ confidence level. The only star that falls out of the $[-3,+3]$ range is ED2-23 (Tab.\ref{tab:abundances1}) which is the Gaia BH3's companion (\citealt{Balbinot+24}, \citealt{Broeck+26}). Hence, the inconsistency can be easily accounted for by its binary nature.\\

\section{Chemical abundance determination}\label{methods}

\subsection{Line selection}
For each star, we identified a list of unblended spectral lines suitable to derive reliable chemical abundances. These lines were selected  using the code K2 (Mucciarelli et al. in prep) according to the line strength of each transition that contribute to an observed feature and taking into account the line broadening  and the signal-to-noise (S/N) ratio of the spectra. Moreover, we calculated the synthetic spectra using the \texttt{SYNTHE} code (\citealt{Sbordone+04}, \citealt{Kurucz+05}) and by computing one-dimensional, plane-parallel, local thermodynamic equilibrium (LTE) model atmospheres using \texttt{ATLAS9} code (\citealt{Sbordone+04}, \citealt{Kurucz+05}), and
the new opacity distribution functions of the KOALA database (\citealt{Mucciarelli+26}).\\
\\
The features were selected in order to consider only those unblended or with contamination by other features leading to a negligible impact in abundance (less than 0.05 dex). A visual check of the selected features was performed on the observed spectra.\\

\subsection{Determination of Fe abundance from equivalent width technique}\label{vt}
The EW technique was applied to isolated and unblended spectral lines unaffected by isotopic splitting or hyperfine structure, using the codes \texttt{DAOSPEC} and \texttt{GALA} (see \citealt{daospec} and \citealt{gala}, respectively).\\
\\
\texttt{DAOSPEC} was used to perform a global normalisation of the observed spectra through polynomial fitting and to measure the EWs of the selected spectral lines. The measured EWs were then analysed with \texttt{GALA}, which derives chemical abundances by comparing the observed EWs with theoretical predictions computed from model atmospheres.\\
\\
In addition to abundance determinations, \texttt{GALA} allows the optimisation of stellar atmospheric parameters. In this work, however, $\rm T_{eff}$ and $\log g$ were fixed to the values derived from Gaia DR3 photometry, while only the microturbulent velocity ($v_{\mathrm{t}}$) was optimised spectroscopically. The value of $v_{\mathrm{t}}$ was determined by imposing the absence of trends between Fe abundances and reduced EWs (defined as $\log(\rm EW)/\lambda$). Since strong lines are the most sensitive to microturbulence, this optimisation was performed only for stars with a statistically significant number of lines located in the first part of the saturation regime of the curve of growth.\\
The stars not satisfying this condition were ED2-9, ED2-10, ED2-11, ED2-19, ED2-20 and ED2-21 (Tab.\ref{tab:abundances1}). Of these six target stars, ED2-9, ED2-10 and ED2-11 are in common with the work of \citealt{Ceccarelli+24} so we decided to fix $v_{\mathrm{t}}$ to the value that they found (0.9 km/s, 0.5 km/s and 0.5 km/s). For ED2-19, ED2-20 and ED2-21 instead we fixed $v_{\mathrm{t}} = 0.9$ km/s. This value corresponds to the mean $v_{\mathrm{t}}$ obtained by \citealt{Ceccarelli+24} for stars with atmospheric parameters similar to those of our targets. Following the same work, typical uncertainties of $0.2$ km/s and $0.4$ km/s were adopted for optimised and fixed $v_{\mathrm{t}}$, respectively.\\

\subsection{Determination of other abundances}
Several of the analysed spectral lines (e.g. Cr, Ba, Sc, and Mn) are affected by hyperfine and/or isotopic splitting, and the EW method is not suitable to take into account these effects. In these cases, the abundances are derived using the spectral synthesis technique. In addition to this, the Na, Mg, and Ca abundances are derived from very strong features (Na D, Mg b, and the Ca II triplet, respectively), which are intrinsically saturated in their cores. The spectral synthesis technique determines chemical abundances through a direct comparison between the observed spectrum and a grid of synthetic spectra computed for different abundance values. The abundance of the element of interest is inferred from the best-fit synthetic spectrum.\\
\\
Specifically, in this project we have employed the code \texttt{SALVADOR} (Alvarez et al. in prep). This code computes on-the-fly synthetic spectra fixing the stellar parameters ($\rm T_{eff}$, $\log g$, $v_{\mathrm{t}}$, [Fe/H]) but varying the abundance of the species of the analysed lines. Finally, through a $\chi^2$ minimisation the code finds the abundance that provides the best match between the observed and the synthetic spectra.\\

\subsection{Uncertainties computation}
Both \texttt{GALA} and \texttt{SALVADOR} provide estimates of the uncertainties associated with the derived chemical abundances by propagating the errors in the adopted atmospheric parameters. In both cases, the uncertainty computation requires as input the variations in $\rm T_{eff}$, $\log g$, and $v_{\mathrm{t}}$ derived in Sect.~\ref{params}. For \texttt{GALA}, after the determination of [Fe/H] and the optimisation of $v_{\mathrm{t}}$, a final run was performed with all atmospheric parameters fixed in order to derive the final Fe abundances and their corresponding uncertainties. In the case of \texttt{SALVADOR}, the atmospheric parameters are fixed by construction during the abundance determination, and the uncertainties are obtained by supplying the corresponding variations in $\rm T_{eff}$, $\log g$ and $v_{\mathrm{t}}$.\\
\\
The total uncertainty on the chemical abundances includes two main contributions: the internal error associated to the measure process (EW or spectral synthesis) and the uncertainty arising from the adopted atmospheric parameters.\\
The internal uncertainty was estimated as the dispersion of the abundances derived from individual lines divided by the square root of the number of lines used. For elements with abundances derived from one single spectral line the internal uncertainty was instead estimated through a Monte Carlo simulation implemented in the \texttt{SALVADOR} code. In this procedure, the best-fit synthetic spectrum is perturbed by adding noise consistent with the S/N ratio of the observed spectrum in the wavelength region around the line of interest, and the abundance determination is repeated 200 times. The standard deviation of the resulting abundance measurements is then adopted as the internal uncertainty. The contribution due to atmospheric parameters, instead, was evaluated by repeating the abundance analysis while varying one parameter at a time by its corresponding uncertainty ($\delta \rm T_{eff}$, $\delta \log g$, and $\delta v_{\mathrm{t}}$), keeping all other parameters fixed. The final uncertainty was then obtained by summing these contributions in quadrature. Therefore the final uncertainties were estimated as follows:
\begin{equation}
    \rm \sigma_{[Fe/H]}=\sqrt{\frac{\sigma_{Fe}^2}{N_{Fe}}+(\delta_{Fe}^{\rm T_{eff}})^2+(\delta_{Fe}^{logg})^2+(\delta_{Fe}^{v_t})^2}
\end{equation}

\begin{equation}\label{ciao}
    \rm \sigma_{[X/Fe]}=\sqrt{\frac{\sigma_{X}^2}{N_{X}}+\frac{\sigma_{Fe}^2}{N_{Fe}}+(\delta_{X,Fe}^{\rm T_{eff}})^2+(\delta_{X,Fe}^{logg})^2+ (\delta_{X,Fe}^{v_t})^2}
\end{equation}
where $\rm \sigma_{X, Fe}$ is the dispersion around the mean of chemical abundances, $\rm N_{X, Fe}$ is the number of used lines, $\delta_{\rm X, Fe}^i$ are the variations in abundances obtained by varying the parameter $i$ ($\rm T_{eff}$, $\log g$ or $v_{\mathrm{t}}$) and $\delta_{\rm X,Fe}^{i}=\delta_{\rm X}^{i}-\delta_{\rm Fe}^{i}$.

\section{Results}\label{results}
\subsection{Metallicity Distribution Function}
\begin{figure}[h]
  \centering
  \includegraphics[width=1.0\columnwidth]{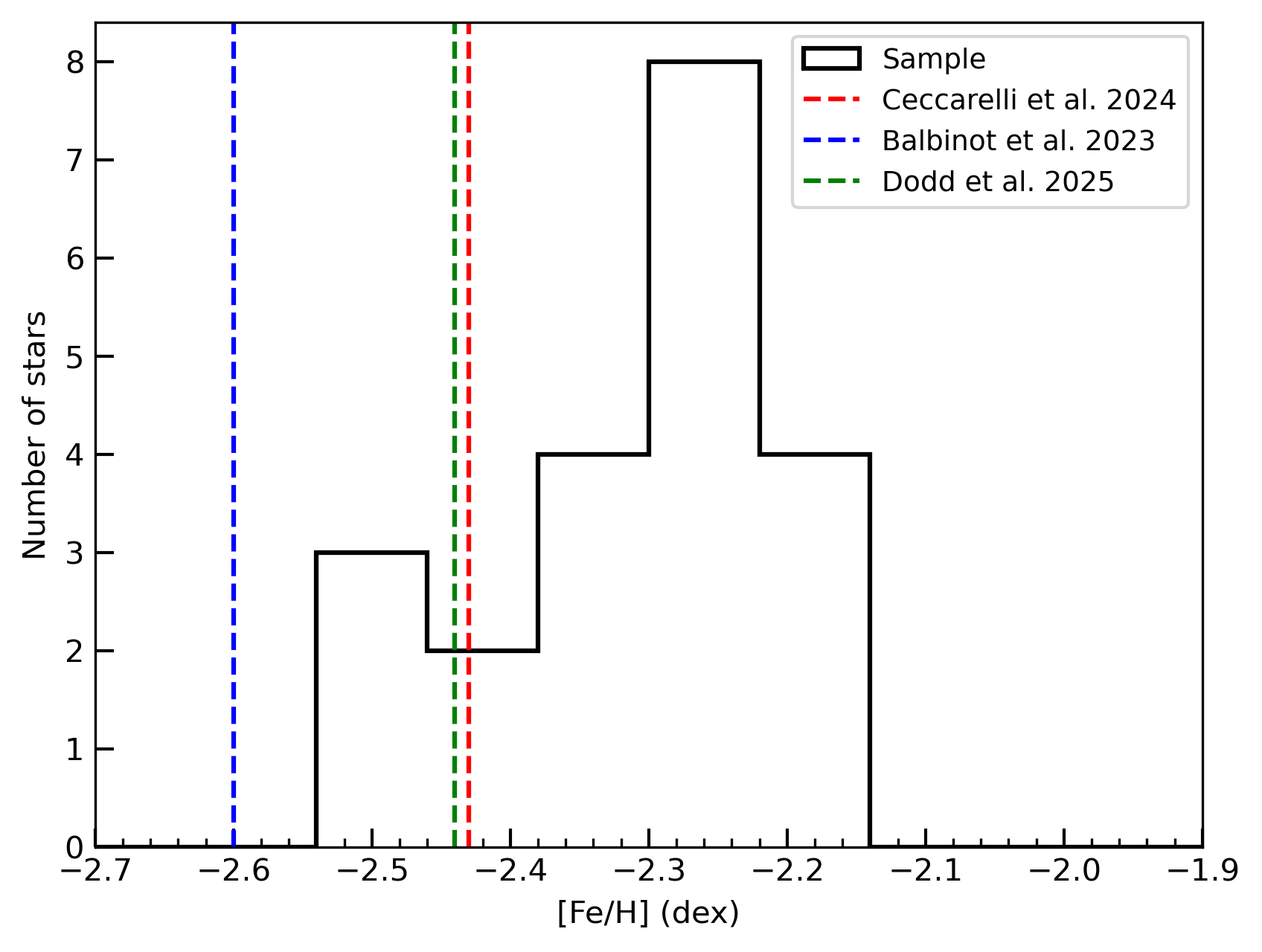}
  \caption{Observed MDF in black. The dashed lines represent the mean value of [Fe/H] found from previous spectroscopic studies of ED-2: in blue the one of \citealt{Balbinot+23}, in red the one of \citealt{Ceccarelli+24} and in green the one of \citealt{Dodd+25}.}
  \label{fig:MDF_obs_full}
\end{figure}

Following previous spectroscopic studies (\citealt{Balbinot+23}, \citealt{Ceccarelli+24}, \citealt{Dodd+25}), we studied ED-2's MDF, which represents the most fundamental chemical difference between the class of systems to which the progenitor of ED-2 could belong. If the progenitor was a stellar cluster, the MDF should be characterised by a negligible intrinsic scatter, $\sigma_{\mathrm{int}}$ (\citealt{Gratton+04}, \citealt{Carretta+09}, \citealt{Bastian_Lardo+18}, \citealt{Carretta+26}). Conversely, dwarf galaxies generally exhibit non-zero intrinsic metallicity spreads as a consequence of their extended or multiple episodes of star formation, which lead to progressive chemical enrichment of the interstellar medium (\citealt{Tolstoy+09}, \citealt{Willman+12}).\\
\\
The MDF derived from the abundance analysis described above is shown in Fig.~\ref{fig:MDF_obs_full}. The observed MDF comprises 21 out of 23 target stars. Specifically, ED2-1 was excluded from our analysis because it is a spectroscopic binary, as indicated by the presence of blended spectral features, which are possibly due to a contamination from another close source. The identification of another spectroscopic binary in ED-2 is particularly interesting given that the system also hosts the black-hole binary Gaia BH3. ED2-18, instead, was excluded because its high effective temperature (7964 K) prevents a reliable abundance determination. ED2-16 was observed both with PEPSI and UVES; in this case, we retained the abundances from the UVES spectrum because its broader wavelength coverage provides access to a larger number of spectral lines, resulting in more robust statistics for the determination of [Fe/H]. Observing the same target star with both UVES and PEPSI, we found an offset of 0.02 dex in the derived [Fe/H] values between the two instruments, consistently with the results of \citealt{Ceccarelli+24}.\\
The mean iron abundance of the sample is [Fe/H] = $-2.32\pm0.03$ dex, with an observed dispersion of $\sigma_{\mathrm{obs}} = 0.15$ dex. For the stars in common with \citealt{Ceccarelli+24} and \citealt{Dodd+25}, we obtain metallicities in agreement with the ones reported in these studies. The small difference between our mean metallicity and those derived by these works therefore arises from the additional stars included in our sample.\\
\begin{figure}[h]
  \centering
  \includegraphics[width=1.0\columnwidth]{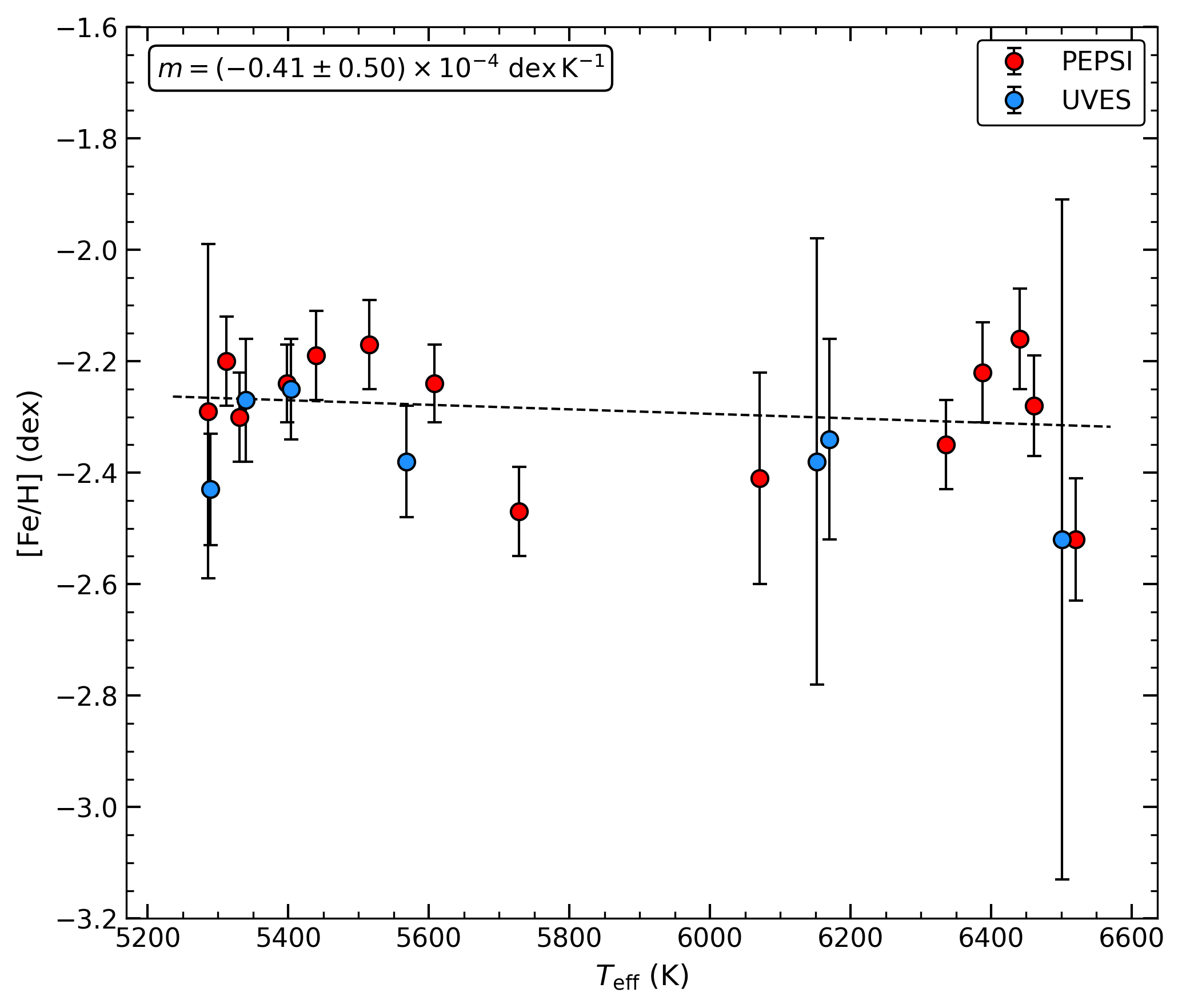}
  \caption{[Fe/H] as a function of $\rm T_{eff}$. In red we have the results for the target stars observed with PEPSI, while in blue those observed with UVES. This result proves that there are no second order effects affecting the metallicity dispersion.}
  \label{fig:Feh_T}
\end{figure}

\begin{figure}[h]
  \centering
  \includegraphics[width=1.0\columnwidth]{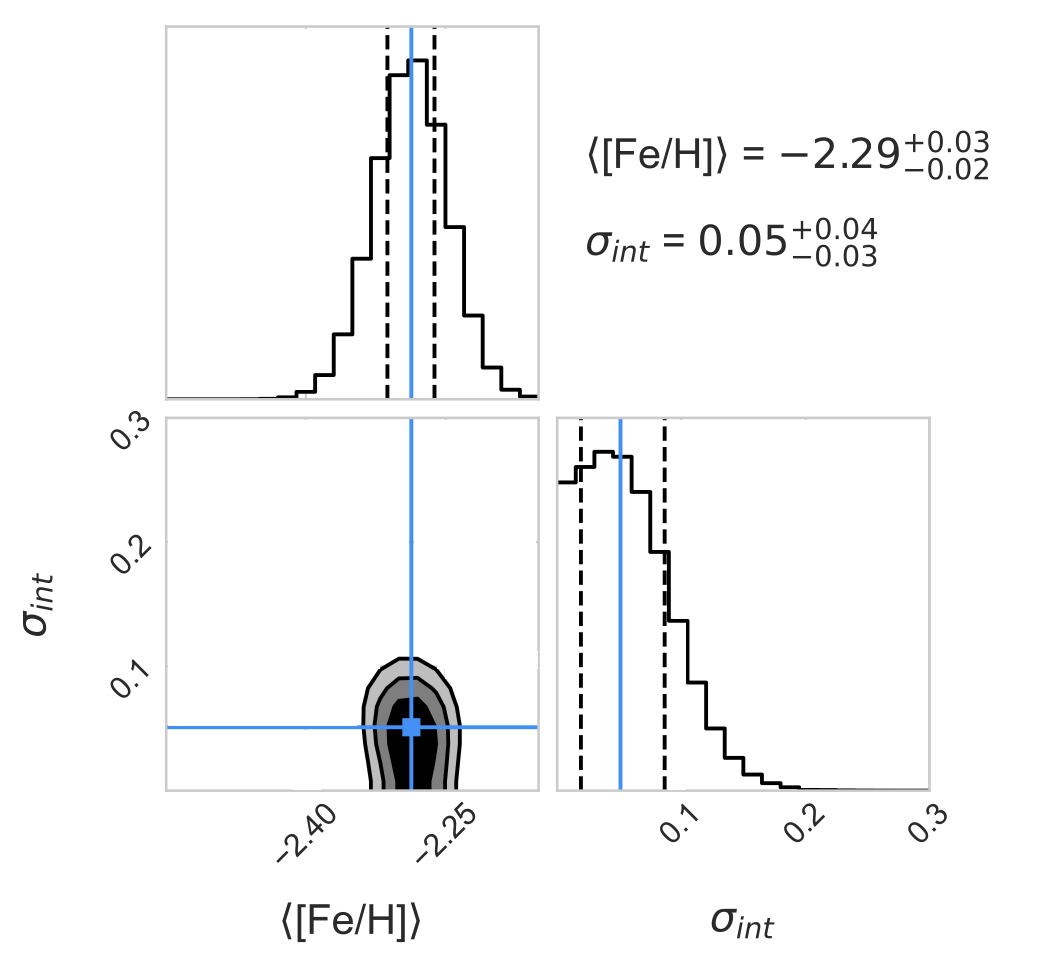}
  \caption{Posterior probability distributions for the mean metallicity and intrinsic metallicity dispersion of the full sample (21 stars), derived through MCMC sampling of the MDF. The diagonal panels show the marginalised posteriors and the 2D posterior distribution function represents the covariance between parameters. We infer [Fe/H] = $-2.29^{+0.03}_{-0.02}$ dex and $\sigma_{\mathrm{int}}=0.05^{+0.04}_{-0.03}$ dex (posterior median and 16th-84th percentile intervals).}
  \label{fig:MDF_MCMC}
\end{figure}

As a first consistency check, we investigated whether the measured [Fe/H] values exhibit any systematic trend with $\rm T_{eff}$ (and therefore with the evolutionary stage), which could artificially broaden the observed metallicity distribution. As shown in Fig.\ref{fig:Feh_T}, no significant correlation between [Fe/H] and $\rm T_{eff}$ is found as the angular coefficient of the best-fitting line is $m=(-0.4\pm0.5)\times 10^{-4}$ dex/K. We therefore conclude that the observed metallicity dispersion is not driven by temperature-dependent systematic effects.\\
\\
To infer the intrinsic metallicity scatter of the MDF, we adopted a Bayesian approach based on Markov Chain Monte Carlo (MCMC) sampling. This method enables a proper treatment of both measurement uncertainties and parameter degeneracies. The observed MDF was modelled as the convolution of an underlying physical metallicity distribution, described by a Gaussian profile characterised by a mean metallicity $\mu$ and an intrinsic dispersion $\sigma_{\mathrm{int}}$, with the individual observational uncertainties properly taken into account. The MCMC analysis returned an intrinsic metallicity dispersion of $\sigma_{\mathrm{int}} = 0.05^{+0.04}_{-0.03}$ dex (Fig.~\ref{fig:MDF_MCMC}), where the quoted uncertainties correspond to the 16th and 84th percentiles of the marginalised posterior distribution. The posterior is consistent with a negligible intrinsic dispersion, with a 95\% upper limit of $\sigma_{\mathrm{int}} < 0.11$ dex.\\

Although the posterior distribution peaks at a small but non-zero value, the intrinsic metallicity dispersion is not formally resolved, as $\sigma_{\mathrm{int}}=0$ is fully consistent with the posterior distribution. Therefore, our analysis provides a strict upper limit on the intrinsic metallicity spread of ED-2.\\
\\
We note that one of the targets, namely ED2-14,  has the largest error in [Fe/H], due to the large uncertainties in the stellar parameters. These large errors arise from the high $\rm E(B-V)$ and the uncertainties in parallax and photometry.\\
Although its metallicity remains consistent within uncertainties with the values derived for the other target stars, we investigated the impact of excluding this object from the MDF analysis. The resulting observed MDF yields a mean metallicity of [Fe/H] = -2.30 dex and an observed dispersion of $\sigma_{\mathrm{obs}} = 0.12$ dex.\\
Applying the same MCMC analysis to this reduced sample, we obtained that both the posterior median metallicity and the inferred intrinsic scatter remain unchanged with respect to the full-sample analysis. Therefore, the MCMC framework already accounts for the large uncertainty associated with the iron abundance determination of this star.

\subsection{Upper limit for the ED2-18's Fe abundance}
\begin{figure}[h]
  \centering
  \includegraphics[width=1.0\columnwidth]{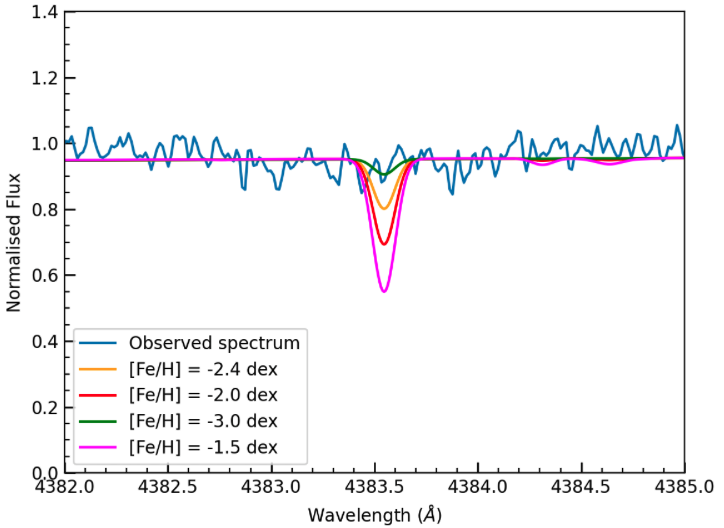}
  \caption{Comparison between the observed spectrum (blue) and a set of synthetic spectra  with different metallicities used in order to provide an upper limit for the Fe abundance of ED2-18. The synthetic spectra are coloured as follows: in green the one with [Fe/H] = -3.0 dex, in orange the one with [Fe/H] = -2.44 dex, in red the one with [Fe/H] = -2.0 dex and in magenta the one with [Fe/H] = -1.5 dex.}
  \label{fig:upper_limit_19}
\end{figure}

ED2-18 (Tab.\ref{tab:abundances1}) is the hottest star in our sample; owing to a spectrum dominated by Balmer lines and telluric features, a reliable abundance analysis could not be performed. Nevertheless, it was possible to derive an upper limit on its Fe abundance.\\ 
To this end, synthetic spectra were computed assuming [Fe/H] = \{-3.0, -2.44, -2.0, -1.5\} dex and compared with the normalised observed spectrum corrected for the $\rm V_{LOS}$ (see Fig.\ref{fig:upper_limit_19}). From this analysis, we inferred an upper limit of [Fe/H] < $-2.0$ dex, fully consistent with the metallicities derived for the remaining stars in our sample.

\subsection{Light elements abundances}\label{light}
\begin{figure*}[h]
  \centering
  \includegraphics[width=1.0\textwidth]{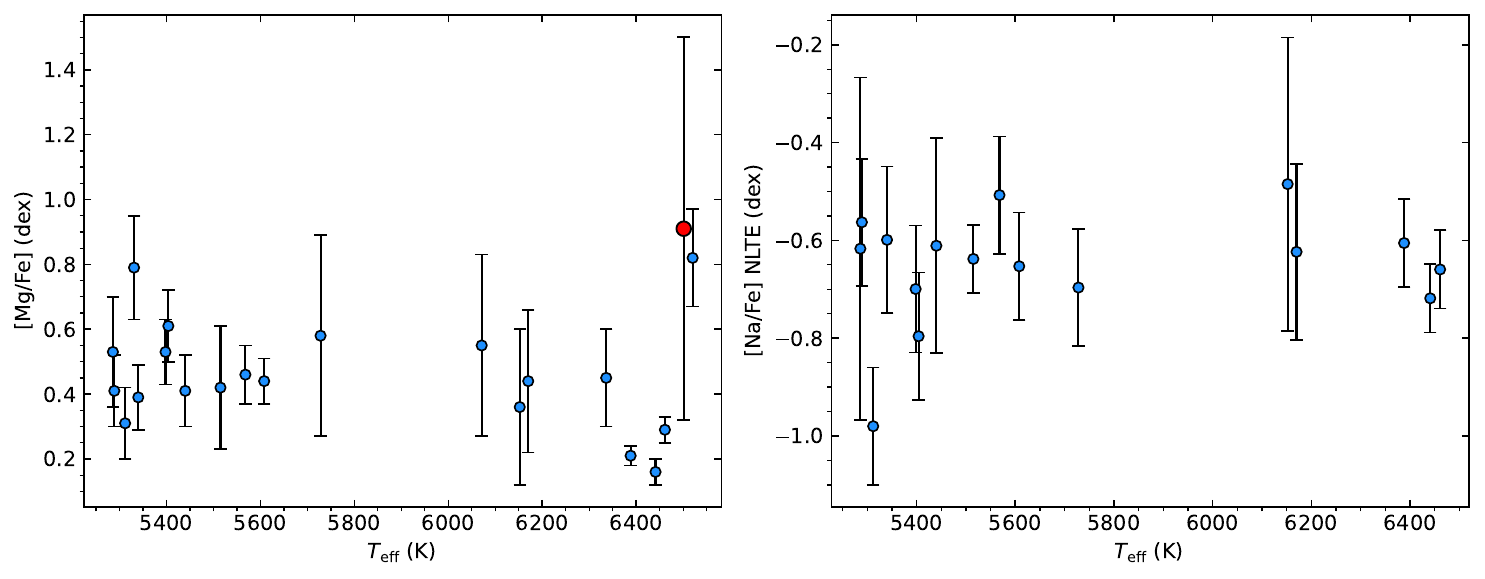}
  \caption{Abundances of Mg and Na shown as a function of effective temperature (left and right panels respectively). In red the red HB star ED2-14}.
  \label{fig:Mg_Na_vs_T}
\end{figure*}

\begin{figure}[h]
  \centering
  \includegraphics[width=0.9\columnwidth]{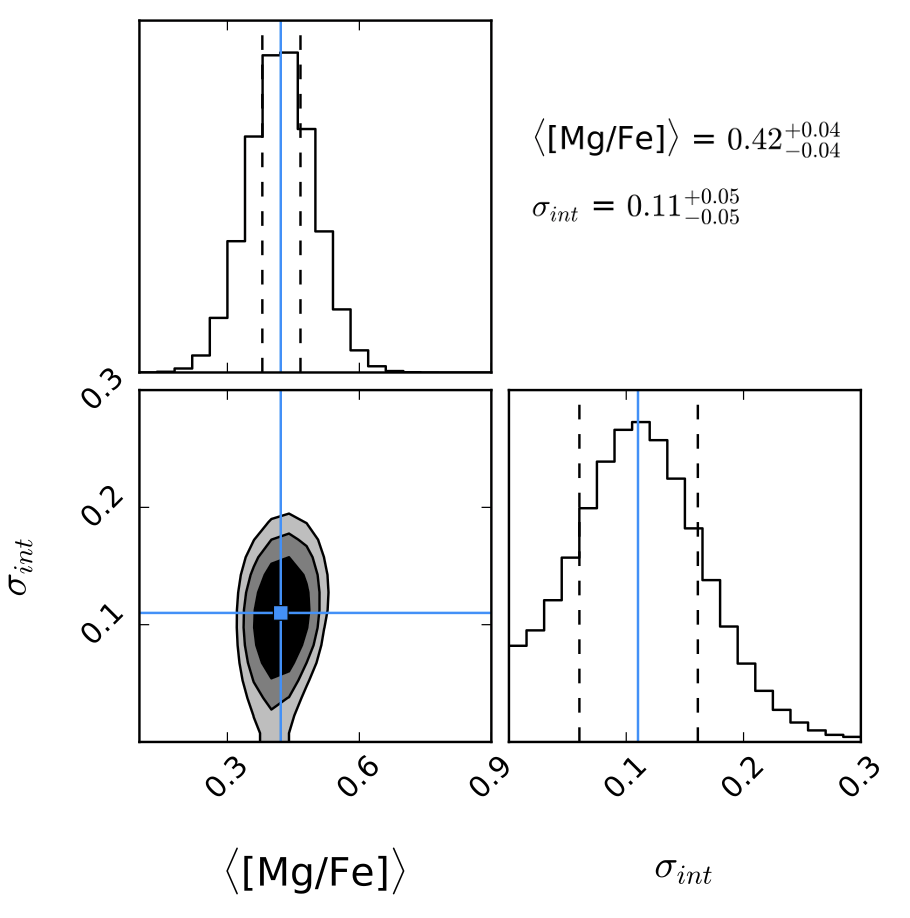}
  \caption{Posterior probability distributions for the mean [Mg/Fe] and intrinsic dispersion, derived through MCMC sampling of the distribution function. The diagonal panels show the marginalised posteriors and the 2D posterior distribution function represents the covariance between parameters. We infer [Mg/Fe] = $0.42^{+0.04}_{-0.03}$ dex, $\sigma_{\mathrm{int}}=0.11^{+0.05}_{-0.05}$ dex (posterior median and 16th-84th percentile intervals).}
  \label{fig:Mg_MCMC}
\end{figure}

\begin{figure}[h]
  \centering
  \includegraphics[width=0.95\columnwidth]{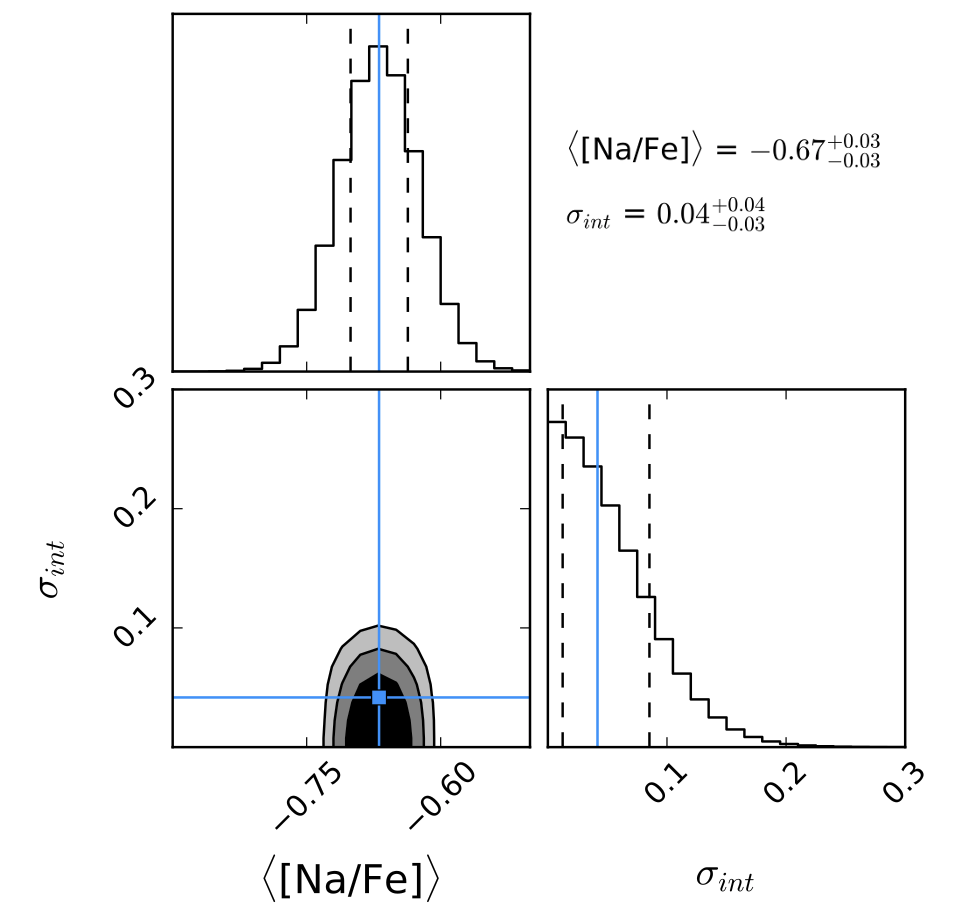}
  \caption{Posterior probability distributions for the mean [Na/Fe] and intrinsic dispersion, derived through MCMC sampling of the distribution function. The diagonal panels show the marginalised posteriors and the 2D posterior distribution function represents the covariance between parameters. We infer [Na/Fe] = $-0.67^{+0.03}_{-0.03}$ dex and $\sigma_{\mathrm{int}}=0.04^{+0.04}_{-0.03}$ dex (posterior median and 16th-84th percentile intervals).}
  \label{fig:Na_MCMC}
\end{figure}

The MDF analysis strongly favours a stellar-cluster origin for ED-2 over a dwarf-galaxy progenitor. Therefore, an important next step is to establish the nature of this stellar cluster and, in particular, to assess whether ED-2 is the remnant of a globular cluster. To this end, we investigated the abundances of light elements, which provide one of the most powerful chemical diagnostics of multiple stellar populations. Indeed, globular clusters are known to exhibit star-to-star abundance variations in light elements such as C, N, O, Na, Mg, and Al, as well as characteristic anti-correlations, including the O–Na and Mg–Al anti-correlations (\citealt{Carretta+09}; \citealt{Bastian_Lardo+18}; \citealt{Gratton+19}). These abundance patterns are not observed in dwarf galaxies or open clusters and are therefore considered a defining signature of globular clusters.\\
\\
Unfortunately, the spectral region containing the diagnostic lines of C, N, and Al is characterised by a low S/N ratio, preventing a reliable abundance determination for these elements. Consequently, the only light elements that could be robustly measured in our sample are Mg and Na. Although these elements alone do not allow a direct search for the canonical anti-correlations, any significant intrinsic abundance spread would provide indirect evidence for the presence of multiple stellar populations. For comparison, metal-poor Galactic globular clusters with [Fe/H] $<-2.0$ dex typically show intrinsic abundance spreads of $\sim$ 0.2 - 0.4 dex in [Na/Fe] and $\sim$ 0.1 - 0.2 dex in [Mg/Fe].\\
For Na, we found that the \texttt{SALVADOR} fits are not sufficiently sensitive to abundance variations because the Na D lines used in the analysis are often saturated. Therefore, we adopted an alternative approach based on the manual measurement of the equivalent widths (EWs) of these lines. Specifically, starting from the continuum-normalised and LOS-velocity-corrected spectra, the EWs were measured through a Gaussian fit of each line over the selected wavelength interval. The measured EWs were then directly used to derive Na abundances including the NLTE corrections provided by the \texttt{INSPECT} database (\citealt{Lind+11}), rather than applying NLTE corrections to LTE abundances obtained from the automatic fitting procedure. This approach allows us to avoid potential uncertainties associated with the fitting of saturated lines and the subsequent conversion of the LTE abundances into the NLTE ones. Moreover, the NLTE corrections are particularly important in this context, as they significantly affect the  intrinsic abundance dispersion: while an intrinsic scatter of $0.11^{+0.05}_{-0.04}$ dex is obtained without NLTE corrections, this decreases to $0.04^{+0.04}_{-0.03}$ dex when they are taken into account. These corrections are reported in Tab.\ref{tab:NLTE_sodium} and were applied to the LTE sodium abundances to obtain the final NLTE [Na/Fe], which are listed in Tab.\ref{tab:abundances2}.\\ 
The derived [Mg/Fe] abundances are shown in the left panel of Fig.\ref{fig:Mg_Na_vs_T}, while the [Na/Fe] abundances, corrected for non-local thermodynamic equilibrium (NLTE) effects using corrections from the \texttt{INSPECT} database, are presented in the right panel. In both cases, the abundances are plotted as a function of $\rm T_{eff}$ (Tab.\ref{tab:abundances1}, Tab.\ref{tab:abundances2}).\\
\\ 
To quantify the intrinsic abundance spreads, we applied the same MCMC framework adopted for the MDF analysis. The resulting posterior distributions are shown in Fig.\ref{fig:Mg_MCMC} and Fig.\ref{fig:Na_MCMC} for Mg and Na, respectively. The MCMC analysis yields [Mg/Fe] $=0.42^{+0.04}_{-0.04}$ dex, $\sigma_{\rm int,Mg}=0.11^{+0.05}_{-0.05}$ dex, [Na/Fe] $=-0.67^{+0.03}_{-0.03}$ dex, and $\sigma_{\rm int,Na}=0.04^{+0.04}_{-0.03}$ dex. Owing to the relatively large abundance uncertainty associated with ED2-14, we repeated the analysis excluding this target from the sample. The resulting posterior distributions were found to be consistent within the uncertainties, indicating that the inferred abundance spreads are not driven by this single star. Overall, the inferred intrinsic dispersions in Mg and Na are not statistically significant.


\subsection{Calcium abundances}
\begin{figure}[h]
  \centering
  \includegraphics[width=1.0\columnwidth]{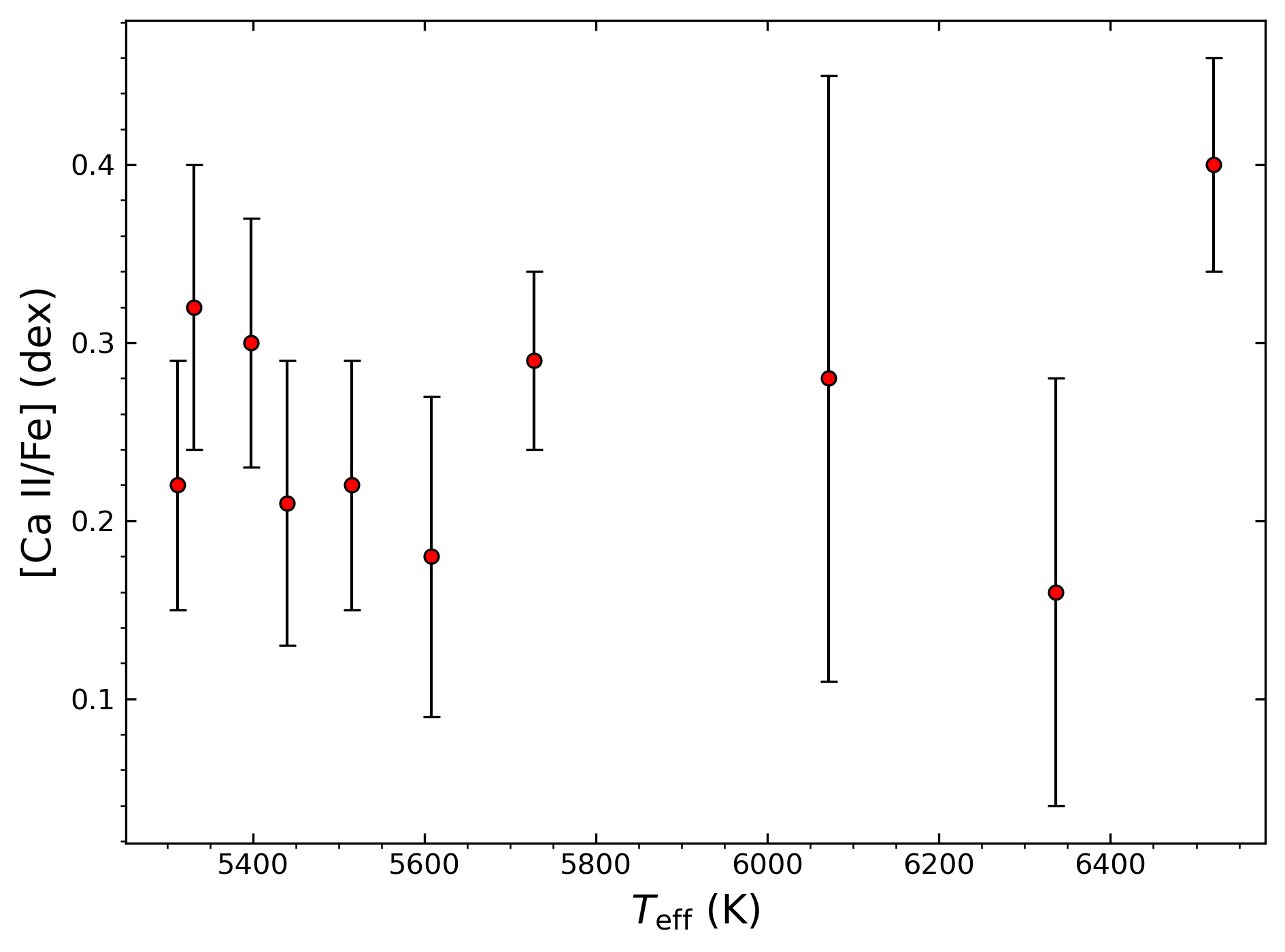}
  \caption{[CaII/Fe] abundance derived from the analysis of Ca II triplet lines at 8498 Å, 8542 Å, and 8662 Å as a function of effective temperature.}
  \label{fig:Ca}
\end{figure}

\begin{figure}[h]
  \centering
  \includegraphics[width=1.0\columnwidth]{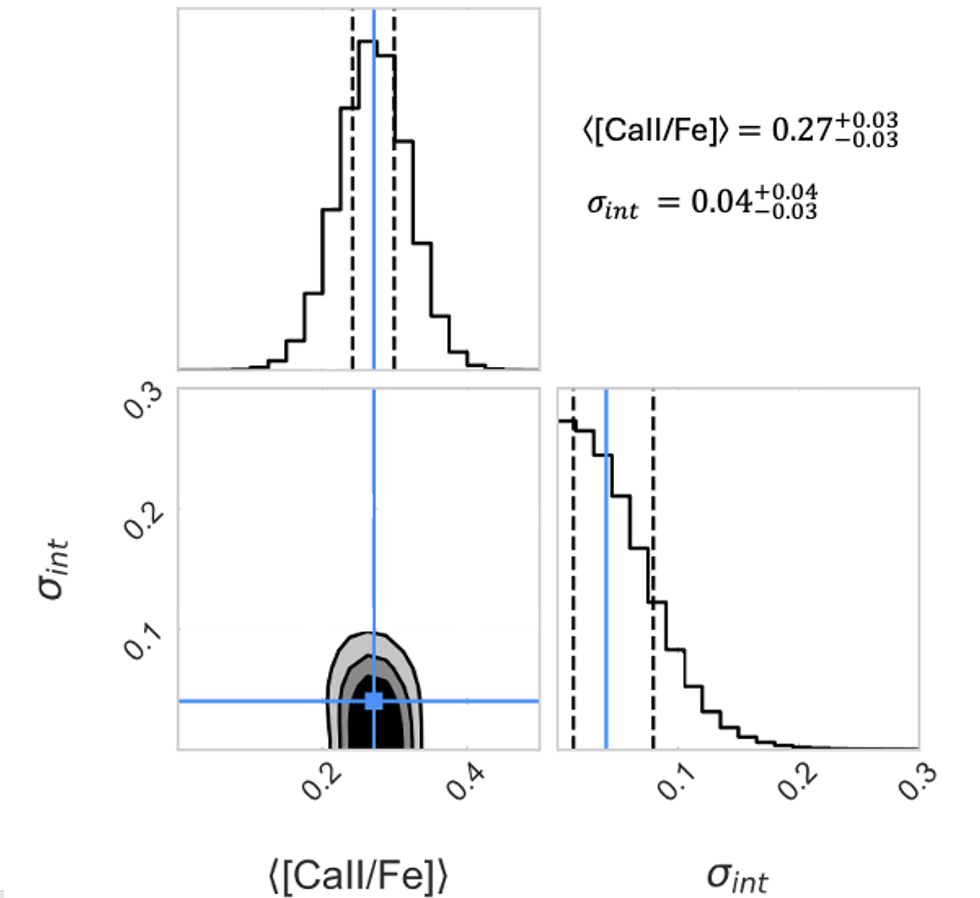}
  \caption{Posterior probability distributions for the mean [CaII/Fe] and intrinsic dispersion, derived through MCMC sampling of the distribution function. The diagonal panels show the marginalised posteriors and the 2D posterior distribution function represents the covariance between parameters. We infer [CaII/Fe] = $0.27^{+0.03}_{-0.03}$ dex and $\sigma_{\mathrm{int}} = 0.04^{+0.04}_{-0.03}$ dex (posterior median and 16th-84th percentile intervals).}
  \label{fig:Ca_MCMC}
\end{figure}

Calcium is an $\alpha$-element, and its abundance was derived primarily from the Ca II infrared triplet lines at 8498 Å, 8542 Å, and 8662 Å whenever these lines were covered by the available spectral range. The abundance derived from neutral calcium (Ca I) lines was used as an independent consistency check on the value inferred from the Ca II triplet. The abundances derived from the two diagnostics were found to be fully consistent within the uncertainties. Therefore, in the following, we focus on the abundances derived from the Ca II triplet, as these are available for the largest fraction of our sample.\\
\\
The resulting [CaII/Fe] abundances are shown in Fig.\ref{fig:Ca}. To quantify the intrinsic abundance spread, we applied the same MCMC framework adopted for the MDF and light-element analyses. The posterior distribution yields a median abundance of [CaII/Fe] = $0.27^{+0.03}_{-0.03}$ dex and an intrinsic scatter of $\sigma_{\mathrm{int}} = 0.04^{+0.04}_{-0.03}$ dex (see Fig.\ref{fig:Ca_MCMC}). The inferred intrinsic dispersion is, therefore, consistent with zero within the uncertainties, indicating the absence of any significant star-to-star variation in Ca abundance among the ED-2 members analysed in this work.
\subsection{Abundances of Ba, Cr, Mn, Ni, K, Sc, Sr and Ti}
\begin{figure*}[h]
  \centering
  \includegraphics[width=0.9\textwidth]{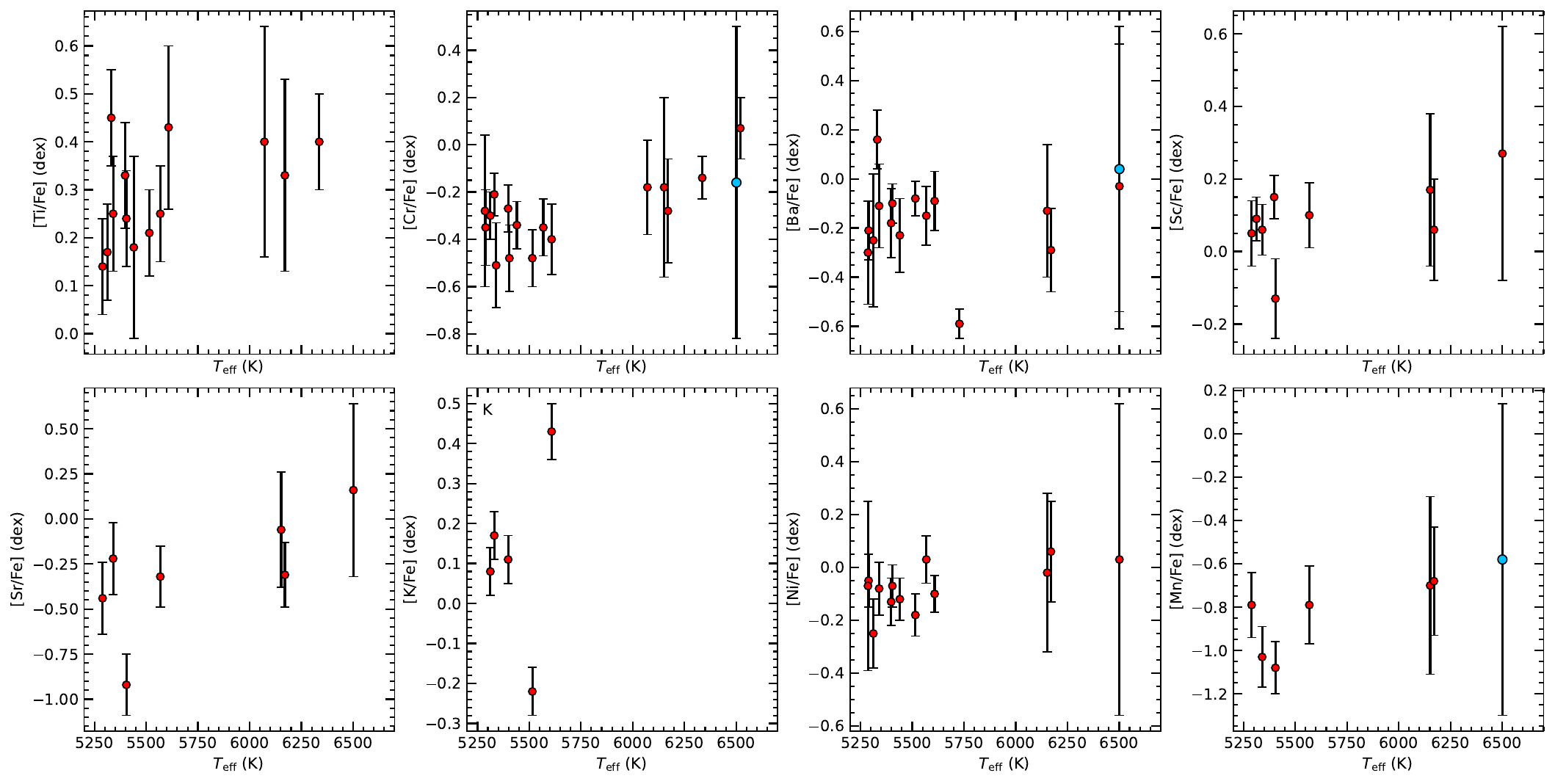}
  \caption{Abundances of Ti, Cr, Ba, Sc, Sr, K, Ni and Mn all represented as a function of effective temperature.}
  \label{fig:abundances}
\end{figure*}

\begin{figure*}[h]
  \centering
  \includegraphics[width=0.9\textwidth]{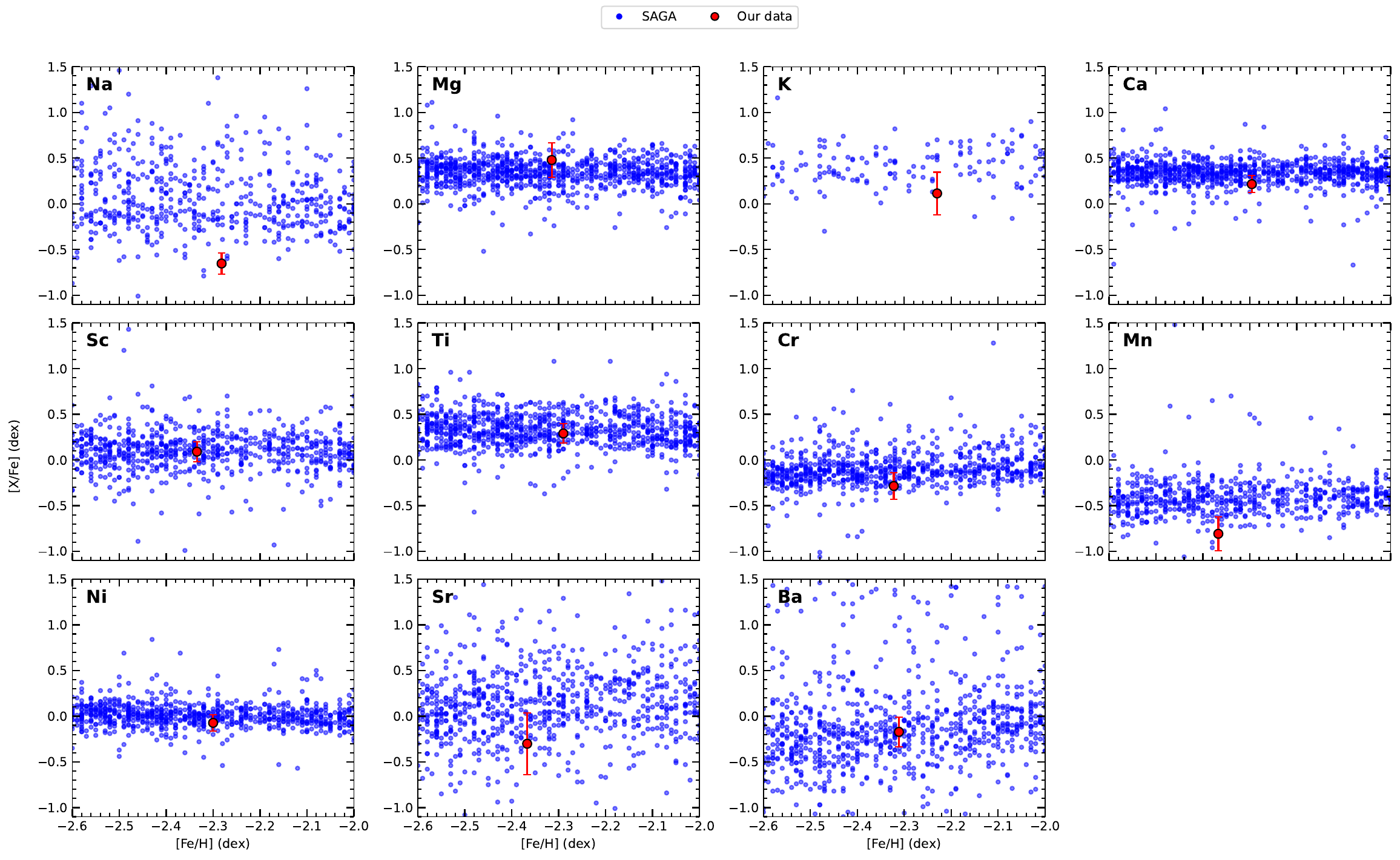}
  \caption{Comparison between the mean abundances found in this work (red points) and the abundances in MW stars taken from the \texttt{SAGA} database (blue points) for Na, Mg, K, Ca, Sc, Ti, Cr, Mn, Ni, Sr and Ba.}
  \label{fig:confronto_saga}
\end{figure*}

To provide the most comprehensive chemical characterisation possible of the stars dynamically associated with ED-2, we also derived the abundances of Ba, Cr, Y, Mn, Ni, K, Sc, Sr, and Ti. The resulting abundance distributions are shown in Fig.~\ref{fig:abundances}, while their mean values and associated uncertainties are reported in Tab.~\ref{tab:abundances2}.\\
\\
For all the measured elements, we calculated the intrinsic abundance spreads using the MCMC analysis discussed above. We found that for all the species, the observed spreads are compatible with a null scatter at a level of about 1$\sigma$ (only for [Sr/Fe] we obtained a spread consistent with a zero within 2$\sigma$). \\
\\
Ba deserves a specific discussion, because only for this element we found a significant intrinsic scatter ($\sigma_{\rm int}=0.18^{+0.05}_{-0.04}$ dex), which is largely driven by the relatively low abundance measured in ED2-3. When this star is excluded from the analysis, the observed scatter is compatible with a null intrinsic dispersion. We note that ED2-3 shows no apparent peculiarity in its atmospheric parameters, no evidence of binary nature (based on the \texttt{ruwe} parameter provided by the Gaia Archive) and no issues with spectral quality that could account for its low Ba abundance, thus supporting the reliability of the measurement. If the spread were true, it would recall the findings of \citealt{Cabrera_Garcia+24} for the metal-poor globular cluster M15, where significant star-to-star variations in neutron-capture elements, including Ba and Eu, were found despite the absence of a significant intrinsic Fe spread. This demonstrates that chemical homogeneity in iron in stellar clusters does not necessarily extend to neutron-capture elements, which may retain signatures of inhomogeneous enrichment due to the stochastic nature of the main sites of rapid neutron capture processes (i.e. neutron star mergers). Although our spectra do not allow us to measure Eu and thus constrain the relative contributions of the r- and s-processes, a similar scenario could apply to ED-2, with the gas from which the system formed being chemically homogeneous in Fe but not necessarily in neutron-capture elements. Further observations aimed at measuring Eu abundances are necessary to investigate whether the abundance pattern observed in M15 is also present in ED-2.\\
\\
In addition to these results, K abundances were determined only for 4 stars, for which the K I line at 7699 \AA\ was clearly detectable and free from telluric contamination. The resulting abundances were corrected for NLTE effects. Specifically, following the approach adopted by \citealt{Alvarez+25}, we applied the NLTE corrections for K from the grid provided by \citealt{Reggiani+19}.
\\
\\
For all the measured abundances, we further assessed whether the abundance ratios measured for ED-2 are consistent with those of MW stars at similar metallicities. To this end, we used the \texttt{SAGA} database (\citealt{SAGAwebsite}) as a reference sample and compared our measurements with the corresponding Galactic abundance trends (Fig.~\ref{fig:confronto_saga}). At these low metallicities, in-situ and accreted stellar populations exhibit largely indistinguishable chemical abundance patterns (e.g. \citealt{Tolstoy+09}); therefore, this comparison is intended primarily as a consistency check of our abundance analysis. While Mg, K, Ca, Sc, Ti, Ni, Sr and Ba are broadly consistent with the \citealt{Kobayashi+20} model, Na, Cr and Mn lie slightly below the predicted trends at [Fe/H] $\approx$ -2.3 dex. Such differences may reflect small variations in the relative contribution of the different nucleosynthetic channels, although uncertainties in the stellar yields and, for some elements, NLTE effects prevent a unique interpretation.\\

\section{Discussion and Conclusions}\label{disc_conc}
As discussed in Sect.\ref{intro}, several studies have investigated the chemical properties of ED-2 in order to constrain the nature of its progenitor system. The first spectroscopic analysis was performed by \citealt{Balbinot+23}, who studied 21 member stars using low and medium-resolution spectra. Although their results pointed towards a narrow MDF, the limited spectral resolution prevented a robust determination of the intrinsic metallicity dispersion. Subsequent studies (\citealt{Balbinot+24}, \citealt{Ceccarelli+24}, \citealt{Dodd+25}, \citealt{Broeck+26}) employed high-resolution spectroscopy, but were restricted to a much smaller number of stars. As a consequence, previous investigations were limited either by spectral resolution or by sample size.\\
\\
The present work overcomes both limitations by combining high-resolution spectroscopy with the largest sample of ED-2 member stars analysed to date. We derive a mean metallicity of [Fe/H] = $-2.29^{+0.03}_{-0.02}$ dex and an intrinsic metallicity dispersion of $\sigma_{\mathrm{int}} = 0.05^{+0.04}_{-0.03}$ dex, consistent with zero within the uncertainties. The mean metallicity derived in this work is broadly consistent with previous high-resolution studies of ED-2 as shown in Fig.\ref{fig:MDF_obs_full}. In particular, the values reported by \citealt{Ceccarelli+24} (-2.43 dex) and \citealt{Dodd+25} (-2.44 dex) agree very well with our determination (see Fig.\ref{fig:MDF_obs_full}) and support the picture of a very metal-poor stellar system characterised by a narrow MDF. An additional validation of our metallicity scale is provided by the independent high-resolution analysis of the black-hole companion system presented by \citealt{Broeck+26}. For this target star (ED2-23), they derived [Fe/H] = $-2.30\pm0.12$ dex, in excellent agreement with our value of [Fe/H] = $-2.29\pm0.30$ dex.\\
\\
The negligible intrinsic dispersion in Fe abundances further strengthens the interpretation of ED-2 as the remnant of a metal-poor stellar cluster accreted into the Milky Way halo, rather than a dwarf galaxy, whose chemical evolution would be expected to produce a measurable metallicity spread.\\
\citealt{Balbinot+24} estimated a rather low mass for this system of $2\times 10^3 \rm M_\odot < M_{ED-2} < 4.2\times 10^4 M_\odot$. To understand whether such a feature is the result of intense mass loss, or whether this cluster was born low mass, we looked into the abundances of light elements. No clear evidence of multiple population signatures is found in our data. We were able to measure Mg and Na abundances, obtaining intrinsic dispersions of $\sigma_{\rm int, Mg}=0.11^{+0.05}_{-0.05}$ dex and $\sigma_{\rm int, Na}=0.04^{+0.04}_{-0.03}$ dex. We take these findings as evidence for the fact that ED-2 was a low mass stellar cluster at birth, and was not able to retain the gas ejected by low-energy sources (see \citealt{Bastian_Lardo+18} for a comprehensive review about the possible polluters in globular clusters).\\
\\
In conclusion, the evidence found in this work suggests that ED-2 might be a new member of an increasingly numerous class of old, metal-poor, and low mass stellar clusters, similarly to the recently discovered C19 (\citealt{Martin+22}) and Koposov 2 (\citealt{Koposov+07}, \citealt{Bellazzini+26}). Future high-quality observations in the blue spectral region will enable a direct search for the abundance patterns involving C, N, and Al that are typically used to identify multiple stellar populations.\\

\begin{acknowledgements}
Based on data acquired using the Large Binocular Telescope (LBT) through the programmes IT-2021B-006 (P.I.: M. Bellazzini), IT-2024B-007 (P.I.: E. Ceccarelli) and the strategic programme IT-2022B-004 (P.I. M. Bellazzini). The LBT is an international collaboration among institutions in the United States, Italy, and Germany. LBT Corporation partners are The University of Arizona on behalf of the Arizona university system; Istituto Nazionale di Astrofisica, Italy; LBT Beteiligungsgesellschaft, Germany, representing the Max-Planck Society, the Astrophysical Institute Potsdam, and Heidelberg University; The Ohio State University; and The Research Corporation, on behalf of The University of Notre Dame, University of Minnesota, and University of Virginia. Based on observations collected at the ESO-VLT under the programmes 0109.B-0522 (P.I. A.: Mucciarelli), 0112.B-0236 (P.I.: E. Ceccarelli) and 0113.B-0196 (P.I.: E. Ceccarelli). This work has made use of data from the European Space Agency (ESA, \href{https://www.cosmos.esa.int/web/gaia}{https://www.cosmos.esa.int/web/gaia}) mission Gaia, processed by the Gaia Data Processing and Analysis Consortium (DPAC, \href{https://www.cosmos.esa.int/web/gaia/dpac/consortium}{https://www.cosmos.esa.int/web/gaia/dpac/consortium}). Funding for the DPAC has been provided by national institutions, in particular the institutions participating in the Gaia Multilateral Agreement. This research has used data, tools or materials developed as part of the EXPLORE project that has received funding from the European Union’s Horizon 2020 research and innovation programme under grant agreement N° 101004214. We thank the anonymous referee for constructive comments that helped us to improve the quality of the manuscript. EC warmly thanks F. Cusano and I. Ilyin for their assistance with the PEPSI data reduction.
\end{acknowledgements}

\bibliographystyle{aa}
\bibliography{biblio}

\appendix
\onecolumn
\section{Summary tables}

\begin{table}[ht!]
\caption{Atmospheric parameters of the ED-2 member stars analysed in this work.}
\label{tab:abundances1}
\centering
\small
\begin{tabular}{lcccccccccccccc}
\hline\hline
Gaia DR3 source ID &  Star &
$T_{\rm eff}$ &
$\log g$ &
$v_{\mathrm{t}}$ &
$\rm V_{LOS}$ &
Spectrum &
Previous analyses &
Notes\\
&
&
(K) &
(dex) &
(km/s) &
(km/s)\\
\hline
3951060094855048960 & ED2-1  & 5664 & 4.40 & ... & -32.2 & PEPSI & Balbinot et al. 2023 & binary\\
3992080193627189632 & ED2-2  & 5515 & 3.44 & 1.6 & -43.7 & PEPSI & Balbinot et al. 2023 & ...\\
759407156314566784 & ED2-3  & 5728 & 3.17 & 1.4 & -150.8 & PEPSI & Balbinot et al. 2023 & ...\\
1693676308288915712 & ED2-4  & 5312 & 3.01 & 1.6 & -461.7 & PEPSI & Balbinot et al. 2023 & ...\\
1577319631286280960 & ED2-5  & 5440 & 3.36 & 1.8 & -356.7 & PEPSI & Balbinot et al. 2023 & ...\\
1578782394069199488 & ED2-6  & 5608 & 3.59 & 1.5 & -372.4 & PEPSI & Balbinot et al. 2023 & ...\\
1698151522476505344 & ED2-7  & 5398 & 3.26 & 1.8 & -456.5 & PEPSI & Balbinot et al. 2023 & ...\\
779616592350301952 & ED2-8  & 5331 & 3.05 & 1.5 & -178.5 & PEPSI & Balbinot et al. 2023 & ...\\
1298276602498467072 & ED2-9  & 6336 & 4.48 & 0.9 & -269.8 & PEPSI & Ceccarelli et al. 2024 & ...\\
4532592619428218624 & ED2-10  & 6071 & 4.47 & 0.5 & -383.4 & PEPSI & Ceccarelli et al. 2024 & ...\\
4245522468554091904 & ED2-11  & 6520 & 4.28 & 0.5 & -304.9 & PEPSI & Ceccarelli et al. 2024 & ...\\
5444820480268295424 & ED2-12  & 5289 & 3.38 & 1.6 & 402.2 & UVES & Balbinot et al. 2023 & ...\\
3549718318990080896 & ED2-13  & 5340 & 3.29 & 1.5 & 332.4 & UVES & Balbinot et al. 2023 & ...\\
5991844282681283712 & ED2-14  & 6501 & 2.91 & 2.7 & 304.0 & UVES& Balbinot et al. 2023 & red HB star\\
6632335060231088896 & ED2-15  & 5568 & 3.58 & 1.7 & 345.4 & UVES & Balbinot et al. 2023 & ...\\
3869876996687740032 & ED2-16  & 5404 & 3.28 & 2.0 & 83.6 & UVES & Balbinot et al. 2023 & observed also with PEPSI\\
4479226310758314496 & ED2-17  & 6152 & 4.43 & 1.5 & -277.2 & UVES & Ceccarelli et al. 2024 & ...\\
4137395670479268352 & ED2-18  & 7964 & 4.25 & ... & -67.2 & UVES & ... & A-star\\
3915925991063287808 & ED2-19  & 6388 & 3.99 & 0.9 & 66.5 & PEPSI & Balbinot et al. 2023 & ...\\
3963873700285600128 & ED2-20  & 6461 & 4.32 & 0.9 & 62.8 & PEPSI & Balbinot et al. 2023 & ... \\
4008274384997266560 & ED2-21  & 6441 & 4.40 & 0.9 & -96.2 & PEPSI & Balbinot et al. 2023 & ...\\
6746114585056265600 & ED2-22  & 6170 & 4.53 & 1.7 & 83.6 & UVES & Dodd et al. 2025 & ...\\
4318465066420528000 & ED2-23 & 5286 & 2.98 & 2.0 & -351.1 & PEPSI & Balbinot et al. 2024 & Gaia BH3's companion\\
\hline
\end{tabular}
\end{table}

\begin{table}[ht!]
\caption{NLTE corrections in dex (taken from \citealt{Lind+11}) for the Na D lines in each target stars.}
\label{tab:NLTE_sodium}
\centering
\small
\begin{tabular}{lcccccccccccccc}
\hline\hline
Star ID & 5889.959 \AA & 5895.913 \AA\\
\hline
ED2-2  & -0.421 & -0.397\\
ED2-3  & -0.282 & -0.143\\
ED2-4  & -0.325 & -0.233 \\
ED2-5  & -0.381 & -0.437 \\
ED2-6  & -0.400 & -0.271 \\
ED2-7  & -0.396 & -0.325 \\
ED2-12  & -0.328 & -0.323 \\
ED2-13  & -0.400 & -0.348 \\
ED2-15  & -0.381 & -0.315 \\
ED2-16  & -0.317 & -0.287 \\
ED2-17  & -0.275 & -0.198 \\
ED2-19  & -0.296 & -0.209 \\
ED2-20  & -0.235 & -0.144 \\
ED2-21  & -0.230 & -0.184 \\
ED2-22  & -0.198 & -0.164 \\
ED2-23 & -0.436 & -0.360 \\
\hline
\end{tabular}
\end{table}

\begin{sidewaystable*} \centering \caption{Chemical abundances for the ED-2 sample.} \label{tab:abundances2} \begin{tabular}{lcccccccccccccc} \hline\hline Star & [Fe/H] & [Mg/Fe] & [Na/Fe] & [Ca/Fe] & [Ba/Fe] & [Cr/Fe] & [Mn/Fe] & [Ni/Fe] & [K/Fe] & [Sc/Fe] & [Sr/Fe] & [Ti/Fe] \\ & (dex) & (dex) & (dex) & (dex) & (dex) & (dex) & (dex) & (dex) & (dex) & (dex) & (dex) & (dex) \\ \hline 
ED2-2 & -2.17$\pm$0.08 & 0.42$\pm$0.18 & -0.64$\pm$0.07 & 0.22$\pm$0.07 & -0.08$\pm$0.07 & -0.48$\pm$0.12 & ... & -0.18$\pm$0.08 & -0.22$\pm$0.06 & ... & ... & 0.21$\pm$0.09 \\ 
ED2-3 & -2.47$\pm$0.08 & 0.58$\pm$0.31 & -0.70$\pm$0.12 & 0.29$\pm$0.05 & -0.59$\pm$0.06 & ... & ... & ... & ... & ... & ... & ... \\ 
ED2-4 & -2.20$\pm$0.08 & 0.31$\pm$0.13 & -0.98$\pm$0.12 & 0.22$\pm$0.07 & -0.25$\pm$0.27 & -0.30$\pm$0.10 & ... & -0.25$\pm$0.13 & 0.08$\pm$0.06 & 0.09$\pm$0.06 & ... & 0.17$\pm$0.10 \\ 
ED2-5 & -2.19$\pm$0.08 & 0.41$\pm$0.13 & -0.61$\pm$0.22 & 0.21$\pm$0.08 & -0.23$\pm$0.15 & -0.34$\pm$0.10 & ... & -0.12$\pm$0.08 & ... & ... & ... & 0.18$\pm$0.19 \\ 
ED2-6 & -2.24$\pm$0.07 & 0.44$\pm$0.12 & -0.65$\pm$0.11 & 0.18$\pm$0.09 & -0.09$\pm$0.12 & -0.40$\pm$0.15 & ... & -0.10$\pm$0.07 & ... & ... & ... & 0.43$\pm$0.17 \\ 
ED2-7 & -2.24$\pm$0.07 & 0.53$\pm$0.13 & -0.70$\pm$0.13 & 0.30$\pm$0.07 & -0.18$\pm$0.14 & -0.27$\pm$0.10 & ... & -0.13$\pm$0.09 & 0.11$\pm$0.06 & 0.15$\pm$0.06 & ... & 0.33$\pm$0.11 \\ 
ED2-8 & -2.30$\pm$0.08 & 0.79$\pm$0.16 & ... & 0.32$\pm$0.08 & 0.16$\pm$0.12 & -0.21$\pm$0.09 & ... & ... & 0.17$\pm$0.06 & ... & ... & 0.45$\pm$0.10 \\ 
ED2-9 & -2.35$\pm$0.08 & 0.45$\pm$0.15 & ... & 0.16$\pm$0.12 & ... & -0.14$\pm$0.09 & ... & ... & ... & ... & ... & 0.40$\pm$0.10 \\ 
ED2-10 & -2.41$\pm$0.19 & 0.55$\pm$0.28 & ... & 0.28$\pm$0.17 & ... & -0.18$\pm$0.20 & ... & ... & ... & ... & ... & 0.40$\pm$0.24 \\ 
ED2-11 & -2.52$\pm$0.11 & 0.82$\pm$0.17 & ... & 0.40$\pm$0.06 & ... & 0.07$\pm$0.13 & ... & ... & ... & ... & ... & ... \\ 
ED2-12 & -2.43$\pm$0.10 & 0.41$\pm$0.15 & -0.56$\pm$0.13 & ... & -0.21$\pm$0.12 & -0.35$\pm$0.16 & -0.79$\pm$0.15 & -0.05$\pm$0.10 & ... & 0.05$\pm$0.09 & -0.44$\pm$0.20 & 0.14$\pm$0.10 \\ 
ED2-13 & -2.27$\pm$0.11 & 0.39$\pm$0.13 & -0.60$\pm$0.15 & ... & -0.11$\pm$0.17 & -0.51$\pm$0.18 & -1.03$\pm$0.14 & -0.08$\pm$0.10 & ... & 0.06$\pm$0.07 & -0.22$\pm$0.20 & 0.25$\pm$0.12 \\ 
ED2-14 & -2.52$\pm$0.61 & 0.91$\pm$0.59 & ... & ... & -0.03$\pm$0.58 & -0.16$\pm$0.66 & -0.58$\pm$0.72 & 0.03$\pm$0.59 & ... & 0.27$\pm$0.35 & 0.16$\pm$0.48 & ... \\ 
ED2-15 & -2.38$\pm$0.10 & 0.46$\pm$0.13 & -0.51$\pm$0.12 & ... & -0.15$\pm$0.12 & -0.35$\pm$0.12 & -0.79$\pm$0.18 & 0.03$\pm$0.09 & ... & 0.10$\pm$0.09 & -0.32$\pm$0.17 & 0.25$\pm$0.10 \\ 
ED2-16 & -2.25$\pm$0.09 & 0.61$\pm$0.15 & -0.80$\pm$0.13 & ... & -0.10$\pm$0.08 & -0.48$\pm$0.14 & -1.08$\pm$0.12 & -0.07$\pm$0.08 & ... & -0.13$\pm$0.11 & -0.92$\pm$0.17 & 0.24$\pm$0.10 \\ 
ED2-17 & -2.38$\pm$0.40 & 0.36$\pm$0.26 & -0.49$\pm$0.30 & ... & -0.13$\pm$0.27 & -0.18$\pm$0.38 & -0.70$\pm$0.41 & -0.02$\pm$0.30 & ... & 0.17$\pm$0.21 & -0.06$\pm$0.32 & ... \\ 
ED2-19 & -2.22$\pm$0.09 & 0.21$\pm$0.10 & -0.61$\pm$0.09 & ... & ... & ... & ... & ... & ... & ... & ... & ... \\ 
ED2-20 & -2.28$\pm$0.09 & 0.29$\pm$0.10 & -0.66$\pm$0.08 & ... & ... & ... & ... & ... & ... & ... & ... & ... \\ 
ED2-21 & -2.16$\pm$0.09 & 0.16$\pm$0.10 & -0.72$\pm$0.07 & ... & ... & ... & ... & ... & ... & ... & ... & ... \\ 
ED2-22 & -2.34$\pm$0.18 & 0.44$\pm$0.23 & -0.62$\pm$0.18 & ... & -0.29$\pm$0.17 & -0.28$\pm$0.22 & -0.68$\pm$0.25 & 0.06$\pm$0.19 & ... & 0.06$\pm$0.14 & -0.31$\pm$0.18 & 0.33$\pm$0.20 \\ 
ED2-23 & -2.29$\pm$0.30 & 0.53$\pm$0.19 & -0.62$\pm$0.35 & ... & -0.30$\pm$0.21 & -0.28$\pm$0.32 & ... & -0.07$\pm$0.32 & ... & ... & ... & ... \\ \hline \end{tabular}

\end{sidewaystable*}

\end{document}